\documentclass[apj]{emulateapj}
\usepackage{comment}
\usepackage{enumerate}
\usepackage[T1]{fontenc}
\usepackage{pslatex}

\shorttitle{FIR AND RADIO MORPHOLOGIES OF GALAXY DISKS}
\shortauthors{MURPHY ET AL.}

\slugcomment{Draft version 2.5 \today}
\journalinfo{Re-submitted to \apj~\today}

\begin{document}

\title{Connecting Far-Infrared and Radio Morphologies of Disk Galaxies:
Cosmic-Ray Electron Diffusion After Star Formation Episodes}

\author{
E.J.~Murphy,\altaffilmark{1} G.~Helou,\altaffilmark{2}
J.D.P.~Kenney,\altaffilmark{1} L.~Armus,\altaffilmark{3} and
R.~Braun\altaffilmark{4}}

\altaffiltext{1}{\scriptsize Department of Astronomy, Yale University,
  P.O. Box 208101, New Haven, CT 06520-8101; murphy@astro.yale.edu}
\altaffiltext{2}{\scriptsize California Institute of Technology, MC
  314-6, Pasadena, CA 91125}
\altaffiltext{3}{\scriptsize {\it Spitzer} Science Center, California
  Institute of Technology, Pasadena, CA 91125} 
\altaffiltext{4}{\scriptsize ASTRON, P.O. Box 2, 7990 AA Dwingeloo,
  The Netherlands}

\begin{abstract}
We present results on the interstellar medium (ISM) properties of 29
galaxies based on a comparison of {\it Spitzer} far-infrared and
Westerbork Synthesis Radio Telescope radio continuum imagery.
Of these 29 galaxies, 18 are close enough to resolve at $\la$1~kpc
scales at 70~$\micron$ and 22~cm.    
We extend the \citet{ejm06a,ejm06b} approach of smoothing infrared
images to approximate cosmic-ray (CR) electron spreading and thus
largely reproduce the appearance of radio images. 

Using a wavelet analysis we decompose each 70~$\micron$ image into one
component containing the star-forming {\it structures} and a second
one for the diffuse {\it disk}.   
The components are smoothed separately, and their combination compared
to a free-free corrected 22~cm radio image; 
the scale-lengths are then varied to best match the radio and smoothed
infrared images.  
We find that late-type spirals having high amounts of ongoing star
formation benefit most from the two-component method.  
We also find that the disk component dominates 
for galaxies having low star formation activity, whereas
the structure component dominates at high star formation activity.  

We propose that this result arises from an age effect rather than from
differences in CR electron diffusion due to varying ISM parameters.
The bulk of the CR electron population in actively star-forming galaxies
is significantly younger than that in less active galaxies due to recent
episodes of enhanced star formation; 
these galaxies are observed within $\sim$10$^{8}$~yr since the onset of the most recent star formation episode.
The sample irregulars have anomalously low best-fit scale-lengths for
their surface brightnesses compared to the rest of the sample spirals
which we attribute to enhanced CR electron escape.    
\end{abstract}

\section{Introduction}
The interstellar medium (ISM) is a complex environment comprised of a
diverse mix of extremely tenuous matter (by terrestrial standards) spanning a 
range of energetic states.    
Atomic, ionized, and molecular material make up a series of gaseous 
({\it thermal}) phases  usually categorized by their temperature and density; 
combinations of these components are used to define the so-called 
two- and three-phase models of the ISM and variations thereof 
\citep[e.g.][and references therein]{cm95}, however, this is not the entire picture.

There is an additional phase of the ISM which is often overlooked due
to the difficulties associated with making direct observations of its
constituents.  
This is the relativistic ({\it non-thermal}) phase 
which is made up of relativistic charged particles known as cosmic
rays (CRs) and magnetic fields. 
The CRs within galaxies have an 
energy density comparable to that of the gaseous phases
\citep[e.g.][]{bc90}.
They fill up the entire volume of galaxies and are important  sources
of heating and ionization of the ISM, though characterizing their
propagation remains an ongoing astrophysical problem  \citep[see][for
  a review]{smp07}.  

The relativistic components of the ISM are important dynamically and
may play a significant role in the regulation of star formation during
the formation and evolution of galaxies \citep[e.g.][and references
  therein]{kf01,cox05}. 
Magnetic fields both help to support interstellar matter against a
galaxy's gravitational potential and confine CRs to galaxy disks; 
thus, magnetic fields and CRs take part in the hydrostatic balance and
stability of the ISM while possibly determining the properties of gas
spiral arms \citep{beck07} and even aiding in the triggering of star
formation \citep{msw74,be82}.  
For sufficiently large CR pressures Parker instabilities \citep{ep66}
can create breaches in magnetic disks, allowing CRs and interstellar
material to freely stream into intergalactic space.

To date, most of our knowledge about the relativistic ISM outside of
the Galaxy has been obtained indirectly through the detection of
synchrotron emission via multi-frequency radio observations
\citep[e.g.][]{nd91,dlg95,ul96,ji99,rb05}.
Synchrotron emission arises from CR electron energy losses as these
particles are accelerated in the magnetic fields of galaxies.
Although the energy density in CR electrons is only $\sim$1\% of that
for CR nuclei, the similarity between the spatial distributions of
gamma-ray and synchrotron emission within the Galaxy suggests that CR
electrons and CR nuclei are fairly well mixed on the scales of a few
hundred parsecs \citep[e.g.][]{ch82,jb86,ww91}.
The spatial distribution of a galaxy's synchrotron emission is a
function of a galaxy's CR electron and magnetic field distributions. 
Thus, radio synchrotron maps provide only limited insight on the
source distribution of the CR electrons as well as the distances the
particles may have traveled before ending up in their current location
of emission.

Massive stars ($\ga$8~$M_{\sun}$) are the progenitors of supernovae
(SNe) whose remnants (SNRs), through the process of diffusive shock
acceleration \citep{ab78,bo78}, appear to be the main acceleration
sites of CR electrons responsible for a galaxy's observed synchrotron
emission. 
These same young massive stars are often the primary sources for dust
heating as they emit photons which are re-radiated at far-infrared
(FIR) wavelengths. 
This shared origin between the FIR and radio emission of galaxies is
thought to be the foundation for the observed FIR-radio correlation
among 
\citep[e.g.][]{de85,gxh85,sn97,nb97,yun01}
and within galaxies 
\citep[e.g.][]{bg88,Xu92,mh95,hoer98,hip03,ejm06a,ah06}.

\begin{deluxetable*}{ccccccccccccc}
  \tablecaption{Basic Galaxy Data\label{tbl-galdat}}
  \tablewidth{0pt}
\tabletypesize{\scriptsize}
  \tablehead{
    \colhead{} & \colhead{R.A.} & \colhead{Decl.} &
    \colhead{$D_{25}$} & \colhead{} & \colhead{} & 
    \colhead{$M_{\rm opt}$} & \colhead{$W_{20}$/$W_{20}^{0}$} & \colhead{$V_{r}$} &
    \colhead{Dist.} & \colhead{$i$} & \colhead{PA} & \colhead{Distance}\\
    \colhead{Galaxy} & \colhead{(J2000)} & \colhead{(J2000)} &
    \colhead{(arcmin)} & \colhead{Type} & \colhead{Nuc.} & 
    \colhead{(mag)} & \colhead{(km s$^{-1}$)} & 
    \colhead{(km s$^{-1}$)} & \colhead{(Mpc)} & \colhead{($\degr$)} &
    \colhead{($\degr$)} & \colhead{References}\\ 
    \colhead{(1)} & \colhead{(2)} & \colhead{(3)} & \colhead{(4)} & 
    \colhead{(5)} & \colhead{(6)} & \colhead{(7)} & \colhead{(8)} &
    \colhead{(9)} & \colhead{(10)}& \colhead{(11)}& \colhead{(12)}& 
    \colhead{(13)}
  }
  \startdata
NGC~628  & 1 36 41.7 & +15 46 59 & 10.5$\times$9.5 & SAc   & \ldots	       & -20.9	& 74	   /175 & 657 & 7.3 &25 & 25 &  1      \\
NGC~925  & 2 27 17.0 & +33 34 43 & 10.5$\times$5.9 & SABcd & H~\textsc{II}     & -20.6	& 224	   /267 & 553 & 9.1 &57 &102 &  2      \\
NGC~2403 & 7 36 51.4 & +65 36 09 & 21.9$\times$12.3& SABcd & H~\textsc{II}     & -19.7	& 257	   /306 & 131 & 3.2 &57 &127 &  2      \\
Holmb~II & 8 19 04.0 & +70 43 09 &  7.9$\times$6.3 & Im    & \ldots	       & -17.1	& 73	   /121 & 157 & 3.4 &37 & 15 &  3      \\
NGC~2841 & 9 22 02.6 & +50 58 35 &  8.1$\times$3.5 & SAb   & Lin/Sy1	       & -20.7	& 611	   /664 & 638 &14.1 &67 &147 &  4      \\
NGC~2976 & 9 47 15.3 & +67 55 00 &  5.9$\times$2.7 & SAc   & H~\textsc{II}     & -17.6	& \ldots/\ldots &   3 & 3.6 &64 &143 &  3      \\
NGC~3031 & 9 55 33.2 & +69 03 55 & 26.9$\times$14.1& SAab  & Lin	       & -21.2	& 446	   /515 & -34 & 3.6 &60 &157 &  2      \\
NGC~3184 &10 18 16.9 & +41 25 28 &  7.4$\times$6.9 & SABcd & H~\textsc{II}     & -19.0	& 142	   /396 & 592 &11.1 &21 &135 &  5      \\
NGC~3198 &10 19 54.9 & +45 32 59 &  8.5$\times$3.3 & SBc   & \ldots	       & -20.2	& 318	   /343 & 663 &13.7 &68 & 35 &  2      \\
IC~2574  &10 28 21.2 & +68 24 43 & 13.2$\times$5.4 & SABm  & \ldots	       & -17.7	& 123	   /134 &  57 & 4.0 &67 & 50 &  6      \\
NGC~3627 &11 20 15.0 & +12 59 30 &  9.1$\times$4.2 & SABb  & Sy2	       & -20.8	& 378	   /417 & 727 & 9.4 &65 &173 &  2      \\
NGC~3938 &11 52 49.5 & +44 07 14 &  5.4$\times$4.9 & SAc   & \ldots	       & -20.1	& 112	   /265 & 809 &13.3 &25 &  0 &  7      \\
NGC~4125 &12 08 05.8 & +65 10 27 &  5.8$\times$3.2 & E6p   & \ldots	       & -21.6	& \ldots/\ldots &1356 &22.9 &58 & 95 &  8      \\
NGC~4236 &12 16 42.1 & +69 27 46 & 21.9$\times$7.2 & SBdm  & \ldots	       & -18.1	& 176	   /185 &   0 & 4.5 &72 &162 &  3      \\
NGC~4254 &12 18 49.5 & +14 24 59 &  5.4$\times$4.7 & SAc   & \ldots	       & -21.6	& 272	   /544 &2407 &16.6 &30 &  0 &  7$^{*}$\\
NGC~4321 &12 22 54.9 & +15 49 21 &  7.4$\times$6.3 & SABbc & Lin	       & -22.1	& 283	   /534 &1571 &14.3 &32 & 30 &  8      \\
NGC~4450 &12 28 29.5 & +17 05 06 &  5.2$\times$3.9 & SAab  & Lin	       & -21.4	& 290	   /433 &1954 &16.6 &42 &175 &  7$^{*}$ \\
NGC~4552 &12 35 39.8 & +12 33 23 &  5.1$\times$4.7 & E     & \ldots	       & -20.8	& \ldots/\ldots & 340 &15.9 &23 &  0 &  9      \\
NGC~4559 &12 35 57.7 & +27 57 36 & 10.7$\times$4.4 & SABcd & H~\textsc{II}     & -21.0	& 251	   /273 & 816 &10.3 &67 &150 &  7      \\
NGC~4569 &12 36 49.8 & +13 09 46 &  9.5$\times$4.4 & SABab & Lin/Sy	       & -22.0	& 360	   /397 &-235 &16.6 &65 & 23 &  7$^{*}$\\
NGC~4631 &12 42 08.0 & +32 32 26 & 15.5$\times$2.7 & SBd   & \ldots	       & -20.6	& 320	   /322 & 606 & 7.7 &83 & 86 &  10     \\
NGC~4725 &12 50 26.6 & +25 30 06 & 10.7$\times$7.6 & SABab & Sy2	       & -22.0	& 410	   /570 &1206 &11.9 &46 & 35 &  2      \\
NGC~4736 &12 50 53.0 & +41 07 14 & 11.2$\times$9.1 & SAab  & Lin	       & -19.9	& 241	   /400 & 308 & 5.0 &37 &105 &  8      \\
NGC~4826 &12 56 43.7 & +21 40 52 & 10.0$\times$5.4 & SAab  & Sy2	       & -20.3	& 311	   /363 & 408 & 5.0 &59 &115 &  7      \\
NGC~5033 &13 13 27.5 & +36 35 38 & 10.7$\times$5.0 & SAc   & Sy2	       & -20.9	& 446	   /501 & 875 &14.8 &63 &170 &  7      \\
NGC~5055 &13 15 49.2 & +42 01 49 & 12.6$\times$7.2 & SAbc  & H~\textsc{II}/Lin & -19.0	& 405	   /489 & 504 & 7.8 &56 &105 &  7      \\
NGC~5194 &13 29 52.7 & +47 11 43 & 11.2$\times$6.9 & SABbc & H~\textsc{II}/Sy2 & -21.4	& 195	   /244 & 463 & 7.8 &53 &163 &  7      \\
NGC~6946 &20 34 52.3 & +60 09 14 & 11.5$\times$9.8 & SABcd & H~\textsc{II}     & -21.3	& 242	   /457 &  48 & 6.8 &32 & 69 &  11     \\
NGC~7331 &22 37 04.1 & +34 24 56 & 10.5$\times$3.7 & SAb   & Lin               & -21.8	& 530	   /561 & 816 &14.5 &71 &171 &  2

\enddata
\tablecomments{Col. (1): ID. Col. (2): The right ascension in the
  J2000.0 epoch. Col. (3): The declination in the J2000.0
  epoch. Col. (4): Major- and minor-axis diameters. Col. (5): RC3
  type. Col. (6): Nuclear type: H~\textsc{II}: H~\textsc{II} region;
  Lin: LINER; Sy: Seyfert (1, 2). Col. (7): Absolute R magnitude, when
  available; otherwise from the V or B bands. Col. (8): Observed/inclination-corrected 21 cm neutral hydrogen line width at 20\% of maximum intensity, 
  in km~s$^{-1}$, taken from \citet{tul88} or RC3.  Col. (9): Heliocentric
  velocity. Col. (10): Distance in Mpc Col. (11): Inclination in
  degrees. Col. (12): Position Angle in degrees. (13): Distance References}

\tablerefs{(1)~\citet{sfe01}; (2)~\citet{wf01}; (3)~\citet{ik02};
  (4)~\citet{lm01}; (5)~\citet{dl02}; (6)~\citet{ik03}; (7)~K. Masters
  2007, in preparation: ($^{*})$ indicates distance set to Virgo
  Cluster center; (8)~\citet{jt01}; (9)~\citet{lf00};
  (10)~\citet{as05}; (11)~\citet{ik00}}
\end{deluxetable*}

Coupling the shared origin of a galaxy's FIR and radio emission
with the fact that the mean free path of dust-heating photons
($\sim$100~pc) is significantly shorter than the expected diffusion
length of CR electrons ($\sim$1-2~kpc) led \citet{bh90} to conjecture
that the radio image of a galaxy should resemble a smoothed version of
its infrared image.  
Consequently, it appears that the close spatial correlation between
the FIR and radio continuum emission within galaxies can be used to
characterize the propagation history of CR electrons.  
This prescription has been shown to hold for galaxies observed at the
``super resolution'' ($\la$1$\arcmin$) of {\it IRAS} HIRES data
\citep{mh98} and, more recently, for high resolution
($\sim$18$\arcsec$) {\it Spitzer} 70$~\micron$ imaging 
\citep[][hereafter M06a]{ejm06a}.
This phenomenology has been further corroborated on
scales $\ga$50~pc by \citet{ah06} who find synchrotron haloes around
individual star-forming regions are more extended than FIR-emitting
regions within the Large Magellanic Cloud.

\citet[][hereafter, M06b]{ejm06b} recently studied how the spatial
distributions of a galaxy's FIR and radio emission vary as a function
of the intensity of star formation. 
They concluded that CR electrons are, on average, younger and closer
to their place of origin within galaxies having higher amounts of star
formation activity compared with more quiescent galaxies.
Using a wavelet-based image decomposition, we extend this work by
attempting to characterize separately
CR electron populations associated with a galaxy's diffuse disk and
its star-forming complexes.  
We carry out this study for a sample of galaxies observed as part of
the {\it Spitzer} Infrared Nearby Galaxies Survey
\citep[SINGS;][]{rk03} and the  Westerbork Synthesis Radio Telescope
\citep[WSRT-SINGS;][]{rb07} for which we have the spatial resolution
to resolve physical scales $<1$~kpc.

The paper is organized as follows:
The galaxy sample is defined in $\S$2 while the
observations and data reduction techniques are discussed in
$\S$3. 
In $\S$4 we present and discuss a correlation analysis between
FIR/radio ratios and various other physical quantities on 
sub-kiloparsec scales within galaxies; 
this section leads us to confirm quantitatively that our
favored phenomenological model is the best description for the
FIR-radio correlation within galaxies.  
In $\S$5 we describe our wavelet-based, two-component image-smearing
analysis; 
the corresponding results are then presented in $\S$6 and their
physical implications are discussed in $\S$7.
We briefly discuss outstanding issues and future prospects in
  $\S$8 and summarize our results and conclusions in $\S$9.

\section{Galaxy Sample \label{sec-samp}}
We present FIR and radio continuum imaging for 29 galaxies included
in the SINGS \citep{rk03} sample.
The SINGS sample consists of 75 {\it normal} galaxies at distances
$\la$30~Mpc. 
These objects were chosen to span a range of Hubble types
(from irregulars to elliptical) as well as exhibit a large range in
star formation rates (SFRs; $<$0.001 to $>$10~$M_{\sun}$yr$^{-1}$),
FIR/optical ratios ($\sim$10$^{3}$), and luminosity
($\sim$10$^{5}$); see \citet{rk03} for a more detailed description of
the SINGS sample. 
The WSRT-SINGS sub-sample of 29 galaxies was then defined by choosing
those SINGS galaxies which could be observed with WSRT (i.e. North of
declination 12$\degr$.5) and have $D_{25} > 5\arcmin$; 
these criteria ensured that the angular resolution across each target would be
acceptable to allow for a resolved study of each galaxy disk.
These 29 galaxies also span a similar range in physical parameters as the parent SINGS sample.
See \citet{rb07} for a more detailed description of the WSRT-SINGS
project.  

In Table \ref{tbl-galdat} we give basic data for each galaxy in the
sample.  
Galaxy diameters ($D_{25}$) and position angles (PA) were taken
from the Third Reference Catalog of Bright Galaxies \citep[RC3;][]{dev91}.
We calculate inclinations using the method described by \citet{dd97}
such that, 
\begin{equation}
\cos^{2}i = \frac{(b/a)^{2} - (b/a)^{2}_{\rm int}}{1-(b/a)^{2}_{\rm
    int}},
\end{equation}
where $a$ and $b$ are the observed semi-major and semi-minor axes and
the disks are oblate spheroids with an intrinsic axial
ratio \((b/a)_{\rm int} \simeq 0.2\) for morphological types earlier
than Sbc and \((b/a)_{\rm int} \simeq 0.13\) otherwise.  
Galaxy distances were taken from the literature with preference given
to direct measurements (see Table \ref{tbl-galdat}).  
For those galaxies where direct measurements were not found we use
flow-corrected estimates (K. Masters 2007, in preparation).
Of these 29 galaxies, 18 have distances less than 11.5~Mpc; these
galaxies are resolved by the 70~$\micron$ {\it Spitzer} beam at spatial
scales $\la$1~kpc.
Due to the high spatial resolution for which we can examine the
FIR-radio correlation within these galaxies, they are the focus of the
present analysis.

\begin{figure*}[!ht]
  \plotone{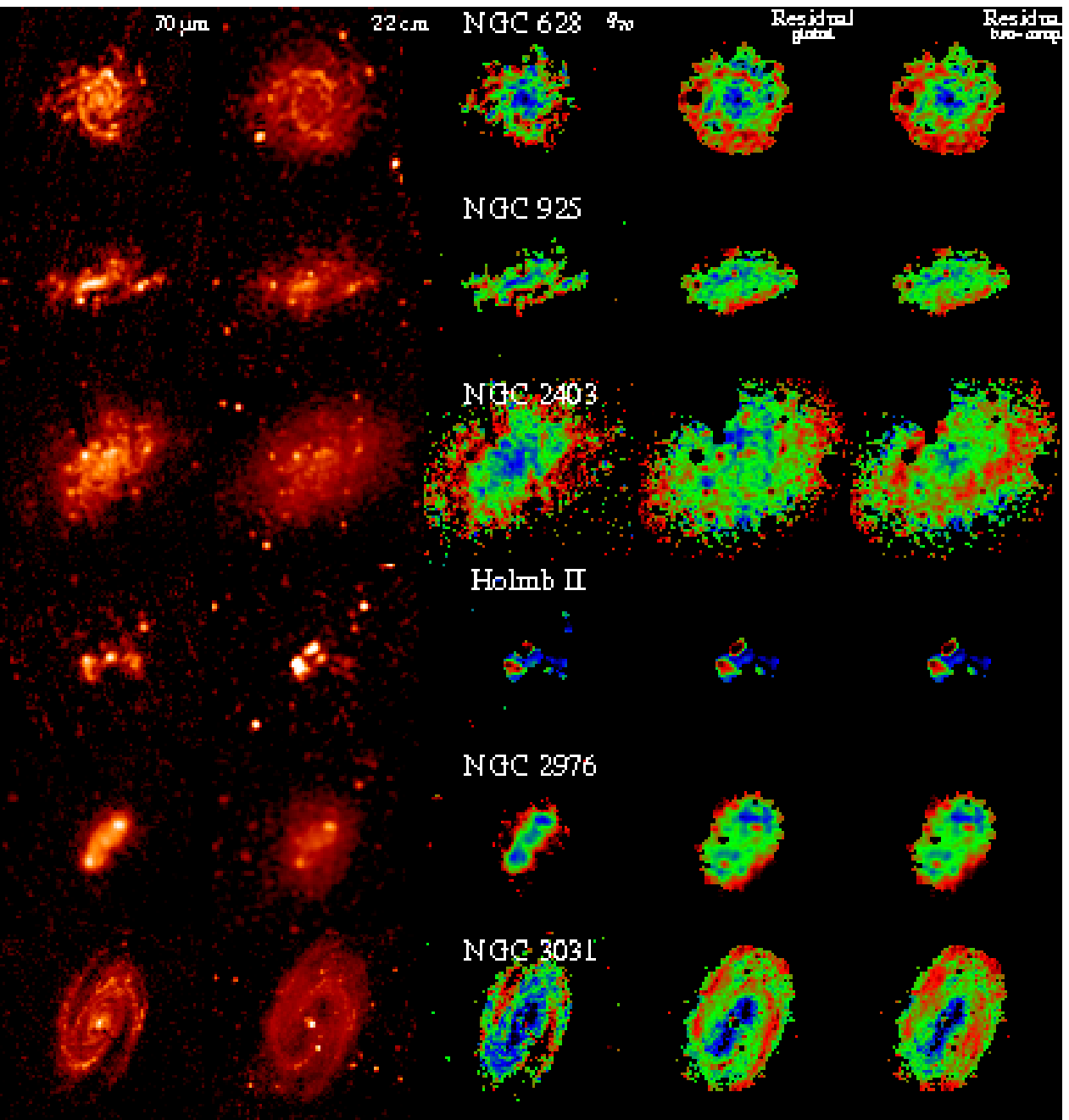}
\caption{
  Each galaxy's 70~$\micron$ and 22~cm images are displayed in columns
  1 and 2, respectively.
  These images are shown using a logarithmic stretch ranging from the
  1-$\sigma$ RMS level of the background to the maximum surface
  brightness of the galaxy.
  A normalized $q_{70}$ map and the residual images between the 22~cm
  and 70~$\micron$ images, smoothed with the best-fit single smearing
  kernel and best-fit disk and structure smearing kernels are given in
  columns 3, 4 and 5 respectively.
  The residual map definition is given in $\S$\ref{sec-smmod}.
  The stretch of each of these 3 maps runs from -0.75 to 0.75~dex; 
  red and blue colors correspond to radio and infrared excesses in the
  residuals, respectively, while green corresponds to residuals
  $\sim$0.  
  Regions removed for the residual calculations (e.g. background
  radio sources) appear as dark, circular holes in a few of the
  residual maps.
  \label{resmaps}}
\end{figure*}

\setcounter{figure}{0}
\begin{figure*}[!ht]
\plotone{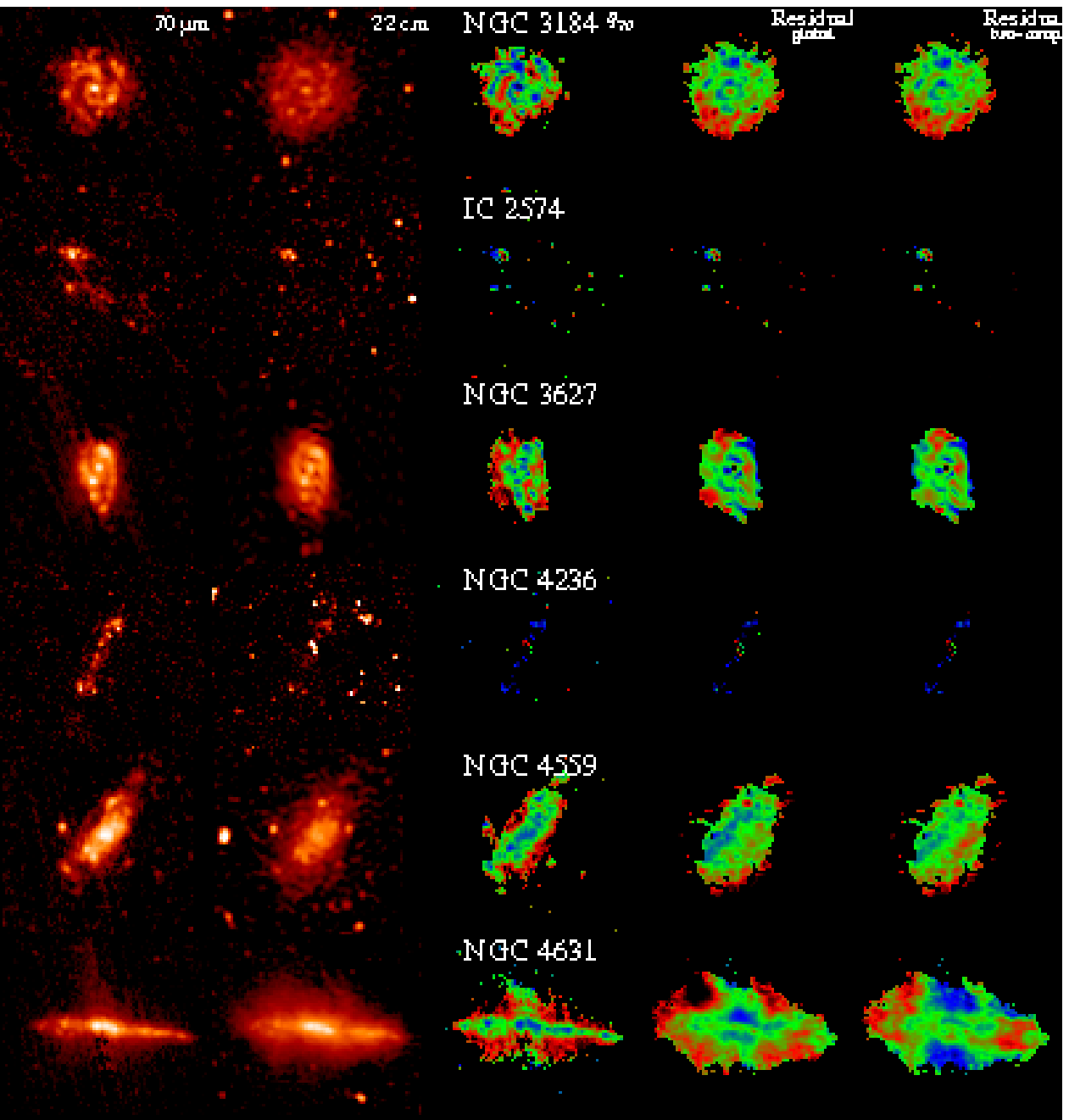}
\caption{\it Continued}
\end{figure*}

\setcounter{figure}{0}
\begin{figure*}[!ht]
\plotone{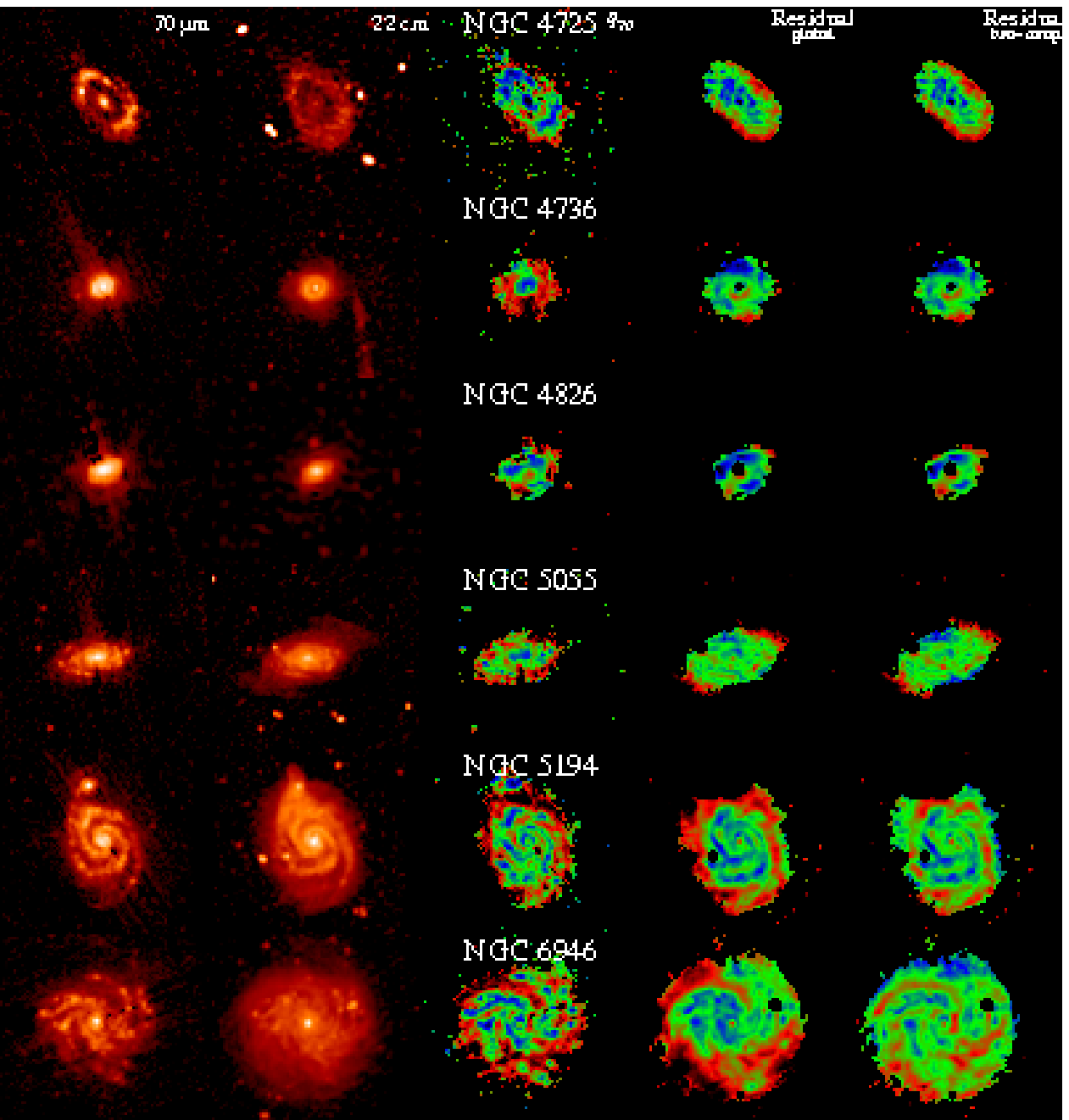}
\caption{\it Continued}
\end{figure*}

\section{Observations and Data Reduction \label{sec-obs}}
\subsection{{\it Spitzer} Images}
Observations at 24, 70, and 160~$\micron$ were obtained using the
Multiband Imaging Photometer for {\it Spitzer} \citep[MIPS;][]{gr04}
as part of the SINGS legacy science program.  
A detailed description of the SINGS observational strategy can be
found in \citet{rk03}.  
The MIPS data were processed using the MIPS Data Analysis Tool
\citep[DAT;][]{kdg05} and included in the SINGS data release 4
(DR4).  
Additional steps beyond the standard reduction procedure of the MIPS
DAT are described in M06a.
The full width at half maximum (FWHM) of the MIPS 24, 70, and
160~$\micron$ point spread functions (PSFs) are 5$\farcs$7, 17$\arcsec$,
and 38$\arcsec$, respectively.
The mean RMS noise values measured for the 24, 70, and 160~$\micron$
maps are $\sim$0.048, 0.447, and 0.574~MJy~sr$^{-1}$, respectively.
SINGS DR4 has improved the final calibration uncertainties to $\sim$4,
$\sim$7, and $\sim$12~\% at 24, 70 and 160~$\micron$, respectively. 
Some artifacts in the MIPS images have also been corrected for in
DR4.
In the 70~$\micron$ data, very high surface brightness regions
are known to be affected by non-linearities in the calibration.
The galaxy most strongly affected is NGC~3627 where only $\sim$5~\%
of the area may be miscalibrated.

\subsection{Radio Continuum Images \label{sec-radobs}}
Radio continuum imaging at 22~cm was performed using the Westerbork
Synthesis Radio Telescope (WSRT) as part of the WSRT-SINGS survey.
A complete description of the radio observations and image processing
steps can be found in \citet{rb07}.  
To mitigate significant flux loss at low spatial frequencies
(i.e. the zero-spacing problem), observations were taken using a
configuration with particularly good sampling at short baselines. 
The WSRT data were CLEANed and self-calibrated using an imaging
pipeline based on the MIRIAD package.  
The final CLEAN maps were restored with 18$\arcsec$ FWHM circular
Gaussian beams.
The intrinsic FWHM of the radio beams is approximately 11$\arcsec$
East-West by 11/sin~($\delta$)$\arcsec$ North-South at 1.4~GHz and
scales as wavelength, where $\delta$ is the source declination.
Total intensity calibration of the data was performed in the AIPS
package and the flux density calibration of the radio maps is better
than 5~\%.
In the case of NGC~3031, which suffers from a combination of low
extended surface brightness and confusion problems due to the nearby
starburst galaxy NGC~3034, a 20~cm map was created via a
variance weighted average of both 22 and 18~cm data in order to
obtain high quality data.
Accordingly, the flux density values of NGC~3031 where scaled to what
is expected at 22~cm assuming a mean spectral index of $-0.8$ to allow
for proper comparison with the rest of the sample.
The mean RMS noise in the radio maps is $\sim$4.4$\times
10^{-3}$~MJy~sr$^{-1}$.

\subsection{Ancillary Data}
Images in the $Ks$-band were taken from the Two Micron All Sky Survey
(2MASS) Large Galaxy Atlas (LGA) \citep{tj03}. 
The typical FWHM of PSF for these images is $\sim$2.5$\arcsec$.

Ultraviolet images were obtained by the Galaxy Evolution
Explorer ({\it GALEX}) at 1528 and 2271~\AA~ as part of the {\it
  GALEX} Atlas of Nearby Galaxies \citep[NGS;][]{gdp07}.  
These galaxies, including nearly all of the SINGS sample, were
observed with relatively deep integration times ($\sim$1.5~ksec). 
With an angular resolution of 4-6$\arcsec$, the {\it GALEX} images are
well matched to {\it Sptizer} 24~$\micron$ imaging and more resolved
than the {\it Sptizer} 70 and 160~$\micron$ data.
For a more detailed description of the {\it GALEX} observations see
\citet{gdp07}.
The only galaxy for which we use {\it GALEX} data that were not obtained as
part of the NGS is NGC~4725; these data were obtained as part of the
All-sky Imaging Survey \citep[AIS;]{cm05} and have much shorter
effective exposure times ($\sim$0.1~ksec). 
Galaxies in the present sample which have not yet been observed by
{\it GALEX}, or for which the data are not publicly available, include
NGC~3184 and NGC~6946.

\begin{deluxetable*}{ccccccccc}
  \tablecaption{Global Flux Densities and Derived Parameters
  \label{globflux}} 
  \tablewidth{0pt}
\tabletypesize{\scriptsize}
  \tablehead{
    \colhead{} &  \colhead{$S_{\nu}$(22~cm)} &
     \colhead{$f_{\nu}$(24~$\micron$)}  &
     \colhead{$f_{\nu}$(70~$\micron$)}  &
     \colhead{$f_{\nu}$(160~$\micron$)} &
     \colhead{$F_{\rm FIR}$} &
     \colhead{$F_{\rm TIR}$} &
     \colhead{}& \colhead{}\\
     \colhead{Galaxy} & \colhead{(Jy)} & \colhead{(Jy)} &
     \colhead{(Jy)} & \colhead{(Jy)} &
     \colhead{(10$^{-13}$~W~m$^{-2}$)} &
     \colhead{(10$^{-13}$~W~m$^{-2}$)} & 
     \colhead{$q_{\rm FIR}$} & \colhead{$q_{\rm TIR}$}
  }
  
  \startdata
NGC~628  &   0.200 &	3.19 &   34.78 &  126.2  &   17.36 &   49.39&	 2.36 &     2.82\\
NGC~925  &   0.090 &	0.95 &   14.40 &   43.33 &    6.67 &   17.47&	 2.30 &     2.71\\
NGC~2403 &   0.360 &	5.84 &   86.36 &  245.6  &   39.64 &  101.5 &	 2.47 &     2.88\\
Holmb~II &   0.005 &	0.20 &    3.67 &    4.46 &    1.36 &	2.73&	 2.82 &     3.12\\
NGC~2841 &   0.100 &	0.91 &   10.22 &   62.29 &    6.53 &   20.80&	 2.24 &     2.74\\
NGC~2976 &   0.068 &	1.37 &   20.43 &   52.56 &    9.18 &   22.61&	 2.56 &     2.95\\
NGC~3031 &0.590$^{a}$&	5.09 &   85.18 &  360.0  &   41.83 &  128.5 &	 2.28 &     2.77\\
NGC~3184 &   0.080 &	1.43 &   15.76 &   70.48 &    8.10 &   25.69&	 2.43 &     2.93\\
NGC~3198 &   0.049 &	1.06 &   10.27 &   39.00 &    5.24 &   15.24&	 2.46 &     2.92\\
 IC~2574 &   0.014 &	0.28 &    5.55 &   11.75 &    2.32 &	5.33&	 2.65 &     3.01\\
NGC~3627 &   0.500 &	7.42 &   92.63 &  230.2  &   42.24 &  102.8 &	 2.35 &     2.74\\
NGC~3938 &   0.080 &	1.09 &   14.25 &   51.98 &    6.97 &   19.87&	 2.37 &     2.82\\
NGC~4125 &   0.002 &   0.079 &    1.11 &    1.77 &    0.46 &	0.96&	 2.81 &     3.13\\
NGC~4236 &   0.026 &	0.55 &    8.27 &   20.43 &    3.67 &	8.92&	 2.58 &     2.96\\
NGC~4254 &   0.510 &	4.20 &   50.29 &  142.9  &   23.69 &   60.65&	 2.09 &     2.50\\
NGC~4321 &   0.310 &	3.34 &   40.59 &  139.6  &   19.79 &   54.95&	 2.23 &     2.67\\
NGC~4450 &   0.013 &	0.21 &    3.42 &   16.94 &    1.82 &	5.79&	 2.57 &     3.08\\
NGC~4552 &   0.093 &   0.094 &    0.54 &    1.42 &    0.29 &	0.72&	 0.92 &     1.31\\
NGC~4559 &   0.110 &	1.12 &   16.89 &   54.15 &    7.93 &   21.35&	 2.28 &     2.71\\
NGC~4569 &   0.170 &	1.44 &   12.37 &   41.21 &    6.29 &   17.23&	 1.99 &     2.43\\
NGC~4631 &   1.290 &	8.15 &  130.2  &  289.5  &   56.22 &  131.5 &	 2.06 &     2.43\\
NGC~4725 &   0.100 &	0.86 &    8.85 &   59.91 &    6.17 &   19.66&	 2.22 &     2.72\\
NGC~4736 &   0.320 &	5.65 &   93.93 &  177.4  &   38.94 &   86.47&	 2.51 &     2.86\\
NGC~4826 &   0.110 &	2.72 &   55.16 &   98.82 &   22.06 &   48.28&	 2.73 &     3.07\\
NGC~5033 &   0.240 &	1.97 &   28.81 &   91.07 &   13.54 &   36.21&	 2.18 &     2.60\\
NGC~5055 &   0.450 &	5.73 &   72.57 &  302.3  &   36.43 &  111.1 &	 2.33 &     2.82\\
NGC~5194 &1.240$^{b}$& 12.67 &  147.1  &  494.8  &   71.84 &  197.5 &	 2.19 &     2.63\\
NGC~6946 &   1.700 &   20.37 &  207.2  &  502.8  &   97.10 &  234.1 &	 2.18 &     2.56\\
NGC~7331 &   0.590 &	4.36 &   74.97 &  189.5  &   33.01 &   80.79&	 2.17 &     2.56
\enddata

\tablerefs{Radio flux densities taken from \citet{rb07}; MIPS flux
  densities taken from \citet{dd07}.}
\tablecomments{$^{a}$: This flux density, measured using a 20~cm map,
  was scaled to the expected flux density at 22~cm assuming a spectral
  index of $-0.8$. 
  $^{b}$: Corrected for flux contributions from companion galaxy
  NGC~5195.}
\end{deluxetable*}

\subsection{Image Registration and Resolution Matching
  \label{sec-resmatch}}  
We match the resolution of the MIPS and radio images using Gaussian
PSFs rather than the MIPS PSFs, which suffer from significant power in
their side-lobes. 
After cropping each set of galaxy images to a common field of view and
re-gridding them to a common pixel scale, we divide the Fourier
Transform (FT) of the final MIPS images at each band by the FT of a
model of the corresponding MIPS PSF.
Next, this quotient is multiplied by the FT of a Gaussian PSF matching
that of the CLEAN beam used to restore 
the corresponding radio images.
The final product was then taken back into the image plane where
we checked to ensure that flux was conserved throughout the resolution
matching procedure. 
The resolution matched 70~$\micron$ and 22~cm maps for each galaxy are
presented in first and second columns of Figure \ref{resmaps},

The 2MASS $Ks$-band images were, likewise, registered, convolved, and
re-gridded to match the MIPS 70~$\micron$ data. 
The same was done for the {\it GALEX} far-UV (1528~\AA; FUV) and near-UV
(2271~\AA; NUV) images after each was first corrected for Milky Way
extinction using the Galactic color excesses of \citet{ds98}. 
The FUV and NUV images were then added to create a single {\it GALEX}
``total UV'' image for each galaxy which we have used in the present
analysis.  

Once the PSF matching of the registered 70~$\micron$ and 22~cm maps
was completed, $q_{70}$ maps were constructed where,
\begin{equation}
  q_{70} \equiv \log~\left[\frac{f_{\nu}(70~\micron)[{\rm Jy}]}
  {S_{\nu}(22~{\rm cm})[{\rm Jy}]}\right].
\end{equation}

\begin{deluxetable}{cccc}
  \tablecaption{Radiation Field Energy Densities and Star Formation
  Rate Surface Densities \label{sfrqdat}}
  \tablewidth{0pt}
  \tablehead{
    \colhead{} & \colhead{$D_{\rm TIR}$} & 
    \colhead{$\log~\Sigma_{\rm TIR}$} &
    \colhead{$\log~U_{\rm rad}$} (TIR+UV)\\
    \colhead{Galaxy} & \colhead{(kpc)} & 
    \colhead{($L_{\sun}$~kpc$^{-2}$)} &
    \colhead{(erg~cm$^{-3}$)}
    }
  \startdata
 NGC~628   &   20.42 &    7.37 &  -12.67\\ 
 NGC~925   &   24.42 &    6.94 &  -13.06\\ 
 NGC~2403  &   12.79 &    7.38 &  -12.60\\ 
 Holmb~II  &	4.76 &    6.57 &  -12.91\\ 
 NGC~2841  &   30.69 &    7.21 &  -12.86\\ 
 NGC~2976  &	5.98 &    7.48 &  -12.38\\ 
 NGC~3031  &   20.12 &    7.19 &  -12.82\\ 
 NGC~3184  &   21.33 &    7.42 &  -12.63\\ 
 NGC~3198  &   27.07 &    7.16 &  -12.89\\ 
 IC~2574   &   10.68 &    6.34 &  -13.32\\ 
 NGC~3627  &   23.10 &    7.80 &  -12.30\\ 
 NGC~3938  &   20.85 &    7.48 &  -12.57\\ 
 NGC~4125  &   13.79 &    6.88 &  -13.00\\ 
 NGC~4236  &   15.50 &    6.33 &  -13.44\\ 
 NGC~4254  &   31.47 &    7.80 &  -12.31\\ 
 NGC~4321  &   30.07 &    7.67 &  -12.43\\ 
 NGC~4450  &   20.17 &    7.11 &  -12.89\\ 
 NGC~4552  &	4.51 &    6.70 &  -12.82\\ 
 NGC~4559  &   24.52 &    7.14 &  -12.89\\ 
 NGC~4569  &   21.50 &    7.59 &  -12.48\\ 
 NGC~4631  &   27.57 &    7.58 &  -12.51\\ 
 NGC~4725  &   30.59 &    7.01 &  -13.03\\ 
 NGC~4736  &   13.76 &    7.61 &  -12.42\\ 
 NGC~4826  &	8.14 &    7.77 &  -12.20\\ 
 NGC~5033  &   40.23 &    7.25 &  -12.84\\ 
 NGC~5055  &   26.45 &    7.56 &  -12.52\\ 
 NGC~5194  &   30.46 &    7.68 &  -12.42\\ 
 NGC~6946  &   27.81 &    7.73 &  -12.38\\ 
 NGC~7331  &   44.81 &    7.47 &  -12.64
 \enddata
\end{deluxetable}

\subsection{Global Flux Densities\label{sec-flux}}
Global flux densities at 22~cm and at each MIPS band were measured by
summing the flux density at each wavelength within elliptical apertures. 
Any identifiable flux contributions from background sources were
masked out in the measurements.  
In the case of NGC~5194, emission from its companion galaxy was also
excluded in the global flux density measurements.
These MIPS and radio flux densities agree with those
presented in \citet{dd07} and \citet{rb07} (within errors). 
We therefore choose to adopt their values for consistency purposes.
The only notable exceptions are the 22~cm flux densities of NGC~3031
and NGC~5194.
Our radio flux density value for NGC~3031 was measured using the
global aperture defined by \citet{dd07} and our averaged 20~cm map
(see $\S$\ref{sec-radobs}).
We then scaled this flux density to what is expected at 22~cm assuming
a mean spectral index of $-0.8$ to allow for proper comparison with
the rest of the sample. 
Unlike \citet{rb07}, we mask out flux contributions from NGC~5195 when
measuring the 22~cm flux density of NGC~5194.
The monochromatic 22~cm, 24, 70, and 160~$\micron$ flux densities for
each sample galaxy are listed in Table \ref{globflux}.

Using a weighted combination of the MIPS flux densities we compute
estimates of the total-infrared (TIR; $3-1100~\micron$) flux of each galaxy 
according to Equation 4 of \citet{dd02}.
We also compute an estimate of the {\it IRAS} based far-infrared (FIR;
$42-122~\micron$) fractions using the same spectral energy
distribution (SED) models of \citet{dd02} such that 
\begin{equation}
\label{firfract}
\frac{F_{\rm FIR}}{F_{\rm TIR}} = \displaystyle\sum_{i=0}^{3}
\xi_{i}\log~\left(\frac{f_{\nu}(70~\micron)}{f_{\nu}(160~\micron)}\right)^{i}, 
\end{equation}
where \([\xi_{0},\xi_{1},\xi_{2},\xi_{3}] = [0.5158, 0.1619, -0.3158,
  -0.1418]\) and \(-0.65 \leq
\log~f_{\nu}(70~\micron)/f_{\nu}(160~\micron) \leq 0.54\).  

The TIR fluxes were then used to calculate global $q_{\rm TIR}$
ratios for each galaxy where,
\begin{equation}
\label{eq-qTIR}
q_{\rm TIR} \equiv \log~\left(\frac{F_{\rm TIR}}{3.75\times10^{12}
  ~{\rm W~m^{-2}}}\right) - \log~\left(\frac{S_{\nu}(22~{\rm cm})}{\rm
  W~m^{-2}~Hz^{-1}}\right).
\end{equation}
The more commonly used $q$ of \citet{gxh85}, which we denote as,
$q_{\rm FIR}$ is also calculated such that 
\begin{equation}
\label{eq-qFIR}
q_{\rm FIR} \equiv \log~\left(\frac{F_{\rm FIR}}{3.75\times10^{12}
  ~{\rm W~m^{-2}}}\right) - \log~\left(\frac{S_{\nu}(22~{\rm cm})}
{\rm W~m^{-2}~Hz^{-1}}\right),
\end{equation}
The FIR and TIR fluxes, along with the associated $q_{\rm FIR}$ and
$q_{\rm TIR}$ ratios, are given in Table \ref{globflux}.

\subsection{Radiation Field Energy Densities \label{sec-sfrd}} 
We compute the radiation field energy density ($U_{\rm rad}$) of
each galaxy using its TIR surface brightness since this parameter is
sensitive to the diffusion of CR electrons.
This was done by first repeating the above image registration and
resolution matching at each MIPS waveband using a 45$\arcsec$ Gaussian
beam (i.e. slightly larger than the FWHM of the native 160~$\micron$
beam) creating TIR surface brightness maps for each galaxy using a
weighted combination of the 3 MIPS band according to Equation 4 of
\citet{dd02} on a pixel by pixel basis.  
The TIR luminosity ($L_{\rm TIR}$) was then summed within elliptical
apertures fit to a galaxy's $1.4\times10^{-7}$~W~m$^{-2}$~sr$^{-1}$
isophotal radius;
this radius corresponds to the maximum 3-$\sigma$ RMS value among the
entire sample's TIR maps and was used to calculate the isophotal
TIR-diameters $D_{\rm TIR}$ for each galaxy.  
We note that these luminosities will be less than those corresponding
to the TIR fluxes given in Table \ref{globflux} as those fluxes were
measured within much larger apertures, effectively going into the
noise of each MIPS map, as given by \citet{dd07}.  
Using the deprojected area of each elliptical aperture, $A_{\rm
  TIR} = \pi(D_{\rm TIR}/2)^{2}$, we calculate TIR surface
brightnesses $(\Sigma_{\rm TIR} = L_{\rm TIR}/A_{\rm TIR})$ along with
estimates of $U_{\rm rad}$ for radiation emitted near the surface
of a semi-transparent body such that
\begin{equation}
\label{eq-urad}
U_{\rm rad} \approx \frac{2\pi}{c}I_{\rm bol} \ga \frac{L_{\rm
    TIR}}{2A_{\rm TIR}c}\left(1 + \sqrt{\frac{3.8 \times
    10^{42}}{L_{\rm TIR}}}\right), 
\end{equation}
where $I_{\rm bol}$ is a galaxy's bolometric surface brightness, $c$
is the speed of light, and all quantities are given in cgs units.
The parenthetical term in Equation \ref{eq-urad} provides a correction
for non-absorbed UV emission that was empirically derived by
\citet{efb03} using archived FIR and UV data for a sample of more than
200 galaxies.  

Since $U_{\rm rad}$ is likely proportional to the disk-averaged
star formation activity within the normal star-forming galaxies
considered here, we also express $U_{\rm rad}$, for illustration
purposes, as a star formation rate surface density, $\Sigma_{\rm
  SFR}$, again using \citet{efb03}.  
The TIR diameters along with values of $\Sigma_{\rm TIR}$ and $U_{\rm
  rad}$, expressed in units of $L_{\sun}$~kpc$^{-2}$ and
  erg~cm$^{-3}$, respectively, are given in Table \ref{sfrqdat}.

\begin{deluxetable*}{ccccccccc}
  \tablecaption{Residual Dispersion Results \label{resdisptbl}}
  \tablewidth{0pt}
  \tablehead{
    \colhead{Galaxy}&
    \colhead{$<q_{70}>$}&
    \colhead{$\sigma_{70}$}&
    \colhead{$\sigma_{\rm C}[I_{\nu}(24~\micron)$]}&
    \colhead{$\sigma_{\rm C}$[Radius]}&
    \colhead{$\sigma_{\rm C}[I_{\nu}(24~\micron)/I_{\nu}(70~\micron)]$}&
    \colhead{$\sigma_{\rm C}[I_{\nu}(Ks)]$}&    
    \colhead{$\sigma_{\rm C}[I_{\nu}({\rm UV})]$}&  
    \colhead{$\sigma_{\rm C}[I_{\nu}(24~\micron)/I_{\nu}({\rm UV})]$}   
  }
\startdata
 NGC~628   &2.26  & 0.258  & 0.166  & 0.189  & 0.258  & 0.183& 0.182  & 0.221\\
 NGC~925   &2.24  & 0.205  & 0.170  & 0.200  & 0.190  & 0.211& 0.177  & 0.196\\
 NGC~2403  &2.27  & 0.253  & 0.133  & 0.171  & 0.248  & 0.165& 0.115  & 0.238\\
 NGC~2976  &2.36  & 0.344  & 0.153  & 0.248  & 0.344  & 0.207& 0.204  & 0.316\\
 NGC~3031  &2.22  & 0.290  & 0.212  & 0.248  & 0.236  & 0.230& 0.256  & 0.259\\
 NGC~3184  &2.22  & 0.226  & 0.173  & 0.204  & 0.202  & 0.195&\nodata &\nodata\\
 NGC~3627  &2.07  & 0.219  & 0.179  & 0.187  & 0.203  & 0.187& 0.188  & 0.209\\
 NGC~4236  &2.31  & 0.454  & 0.395  & 0.452  & 0.447  & 0.515& 0.411  & 0.446\\
 NGC~4559  &2.23  & 0.232  & 0.159  & 0.206  & 0.224  & 0.199& 0.170  & 0.213\\
 NGC~4631  &1.76  & 0.245  & 0.204  & 0.226  & 0.237  & 0.183& 0.221  & 0.242\\
 NGC~4725  &2.24  & 0.266  & 0.155  & 0.247  & 0.237  & 0.216& 0.241  & 0.230\\
 NGC~4736  &2.14  & 0.234  & 0.206  & 0.208  & 0.193  & 0.210& 0.208  & 0.234\\
 NGC~4826  &2.45  & 0.211  & 0.183  & 0.179  & 0.167  & 0.173& 0.169  & 0.209\\
 NGC~5055  &2.06  & 0.244  & 0.196  & 0.214  & 0.168  & 0.213& 0.206  & 0.212\\
 NGC~5194  &1.92  & 0.234  & 0.172  & 0.210  & 0.233  & 0.176& 0.175  & 0.232\\
 NGC~6946  &1.93  & 0.256  & 0.179  & 0.235  & 0.244  & 0.228&\nodata &\nodata\\
\hline \hline
{\bf Averages} & {\bf 2.17}  & {\bf 0.261}  & {\bf 0.190}  & {\bf 0.226}  & {\bf 0.239}  & {\bf 0.218} & {\bf 0.209}  & {\bf 0.247} 
\enddata
\end{deluxetable*}

\section{Correlation Analysis of $\lowercase{q}_{70}$ 
  \label{sec-kpcan1}}
M06a showed that the FIR/radio ratio is not constant within galaxies,
exhibiting strong variation patterns in residual maps and a dispersion
as small as the dispersion in the FIR/radio ratios among galaxies. 
Some of the variation in the FIR/radio ratios within galaxies was
found to be related to both infrared surface brightness and radius.
Using our larger sample, we confirm these results here as well as
expand the explored parameter space.
We search for the dominant physical parameter related to the
variations in the FIR/radio ratio within galaxies by plotting 
$q_{70}$ against a number of physical quantities and calculating the
residual dispersion in each fit.  
This was done for galaxies sufficiently nearby that the resolution at
70~$\micron$ is better than 1~kpc.

\subsection{$q_{70}$-Correlation Analysis within Galaxies
  \label{sec-kpcan2}} 
We examine the relations between $q_{70}$ and: 
a) 24~$\micron$ surface brightness: since it has been found to trace
local massive star formation activity \citep[e.g.][]{dc05, dc07}; 
b) $Ks$-band surface brightness: tracer of stellar mass; 
c) UV surface brightness: extinction dependent tracer of
local star formation; 
d) galactocentric radius;
e) 24/70~$\micron$ surface brightness ratio: local heating of dust;
f) 24~$\micron$/UV surface brightness ratio: obscured star-formation
and permeability of ISM.
Each of the UV, $Ks$, 24~$\micron$, 70~$\micron$, and 22~cm
images were re-gridded with 1~kpc$^{2}$ pixels. 
Pixels in the re-gridded images of each galaxy were removed if they
were either below the 3-$\sigma$ level or associated with emission
from non-related objects, such as background galaxies. 
Galaxies having $<$10 such resolution elements remaining were excluded
in the analysis (i.e. Holmb~II and IC~2574).

\begin{figure*}
\plotone{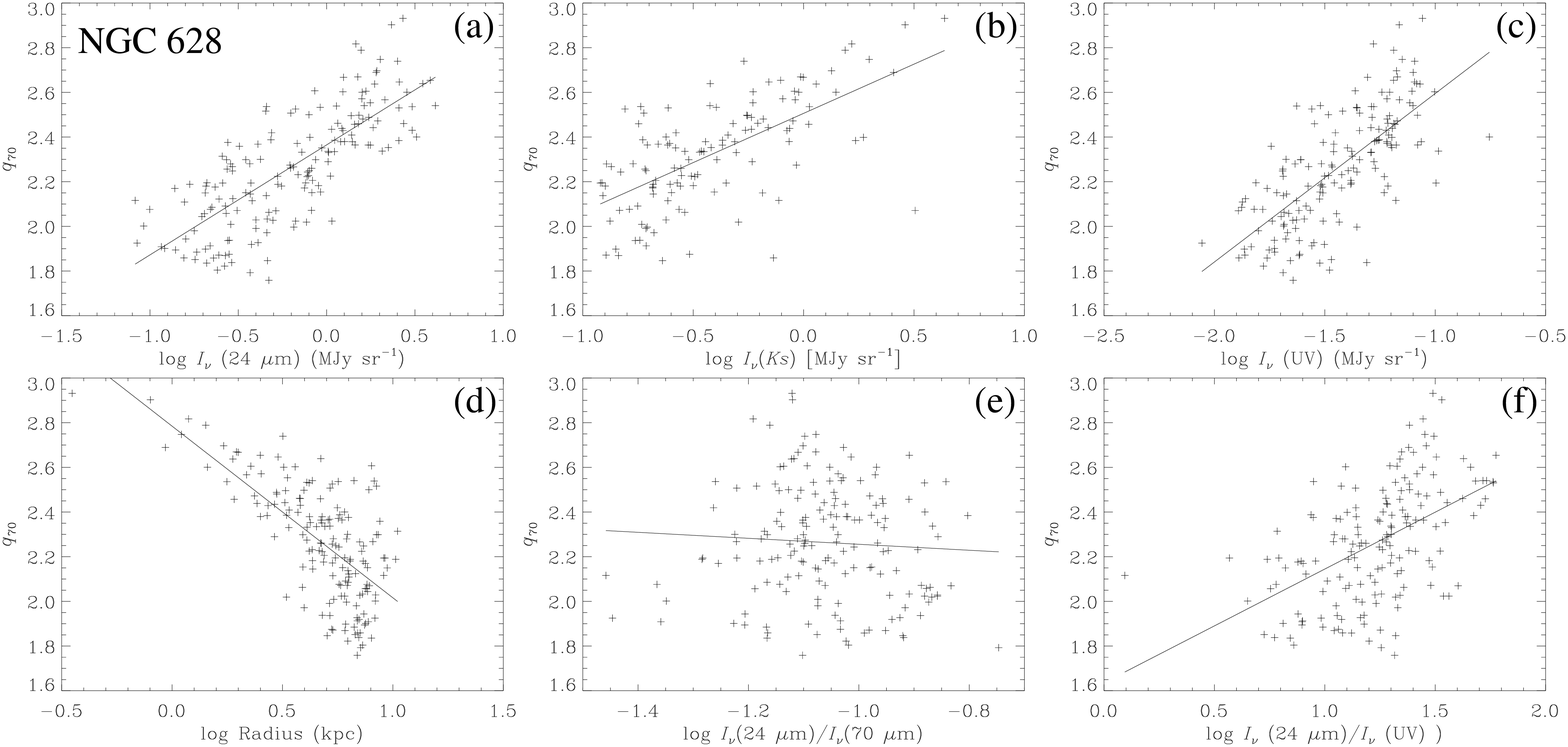}
\caption{Plots comparing the dependence of $q_{70}$ on 24~$\micron$ surface
  brightness, $Ks$ surface brightness, {\it GALEX} UV surface brightness
  (FUV + NUV), galactocentric radius,
  $[I_{\nu}(24~\micron)/I_{\nu}(70~\micron)]$ flux density ratio, and
  $[I_{\nu}(24~\micron)/I_{\nu}({\rm UV})]$ flux density ratio are
  plotted in panels (a), (b), (c), (d), (e), and (f), respectively,
  for NGC~628.   
  Each data point corresponds to a measurement from 1~kpc$^{2}$
  pixels. 
  Ordinary least squares (OLS) fits, which were used to measure the
  residual dispersion, are overplotted in each panel.
  \label{kpcplots}} 
\end{figure*}

An example of the $q_{70}$ scatter-plots for each of the six tested
parameters are given for NGC~628 in Figure \ref{kpcplots}.
Each relation is fit using an ordinary least squares (OLS) regression
of the form \(q_{70} = {\rm m}\log~(X) + {\rm b}\), where $X$ is one
of the six parameters cited above, m is the slope of the fit, and
b is the y-intercept of the fit.
To quantitatively determine which of these six parameters best
correlates with the FIR/radio ratio, we compute the residual
dispersion about the regression line, denoted as $\sigma_{\rm C}[X]$.
Average $q_{70}$ ratios and the associated dispersions are given
in Table \ref{resdisptbl} along with the residual dispersions for
each parameter and galaxy;  
the means for the entire sample are also presented.

\subsection{Results of $q_{70}$ Correlation Analysis
  \label{sec-kpcan3}}  
We find that the residual dispersion around the trend of $q_{70}$
versus 24~$\micron$ surface brightness is, on average, $\sim$0.07~dex
smaller than the dispersion in $q_{70}$;
this is the largest improvement among all the parameters inspected.   
The average values of 
$\sigma_{\rm C}[I_{\nu}(Ks)]$,  $\sigma_{\rm C}[I_{\nu}({\rm UV})]$, 
$\sigma_{\rm C}[\rm Radius]$, 
$\sigma_{\rm C}[I_{\nu}(24~\micron)/I_{\nu}(70~\micron)]$, 
and $\sigma_{\rm C}[I_{\nu}(24~\micron)/I_{\nu}({\rm UV})]$, 
are $\sim$0.03, 0.02, 0.04, 0.05, and 0.05~dex larger,
respectively, than the average value of $\sigma_{\rm
  C}[I_{\nu}(24~\micron)]$ suggesting that the FIR/radio ratio
is most strongly correlated on 24~$\micron$ surface brightness.
A galaxy's 24~$\micron$ surface brightness appears to be a good proxy
for extinction corrected Pa$\alpha$ and therefore the amount of
ongoing massive star formation activity within a galaxy
\citep[e.g.][]{dc05}; 
however, we note that a combination of 24~$\micron$ and H~$\alpha$
emission, to account for unobscured star formation, has been found to
be a better proxy than 24~$\micron$ emission alone
\citep[i.e.][]{dc07, rk07}.  
The fact that the residual dispersion between $q_{70}$ and UV surface
brightness is nearly as small as the residual dispersion around the
trend of $q_{70}$ and 24~$\micron$ surface brightness suggests that,
at least at kpc scales, most of the star formation within our sample
is not highly obscured.
This result is consistent with that of \citet{mp07} who find that
highly-obscured star forming regions make up only $\sim$3\% of a
sample of $\sim$1800 star-forming regions within SINGS galaxies on
scales down to $\sim$500~pc.

Comparing the parameters of radius and $Ks$ surface brightness, we
find that the residual dispersions are similar. 
This result is not too surprising as the $Ks$ surface brightness
samples the older stellar population of a galaxy which, generally, has
a smooth spatial distribution as a function of radius.
We also find that the residual dispersion around the trend of $q_{70}$
versus $I_{\nu}(24~\micron)/I_{\nu}(70~\micron)$ is only slightly
smaller than the dispersion in $q_{70}$, and significantly larger than
the residual dispersion around the trend of $q_{70}$ versus
$I_{\nu}(24~\micron)$ suggesting $q_{\rm 70}$ is not very
sensitive to temperature changes.

\subsection{Local FIR/Radio Ratio and Star Formation}
The above results demonstrate that, at least on kpc scales, the local
FIR-radio correlation is strongly coupled to star-formation activity.
However, from an empirical standpoint, identifying physical parameters
to minimize the residual dispersion in $q_{70}$ is not where the
greatest improvement has been found.
While we find an average decrease in the residual dispersion between
$q_{70}$ and 24~$\micron$ surface brightness of $\sim$20\%, M06b has
shown that a phenomenological smearing model improves the correlation
between the FIR and radio emission distributions of galaxies by a
factor of $\sim$1.8. 
This significantly larger improvement using the image-smearing
technique likely arises because this method more precisely
characterizes the time evolution of a galaxy's FIR and radio emission
distributions, 
which depend critically on its star formation history.
The image-smearing procedure minimizes the residuals between a
galaxy's infrared and radio images after smoothing the infrared map by
a parameterized kernel assumed to contain the diffusion and decay 
characteristics of a galaxy's CR electrons.  
Since CR electrons spread over time, the size of each smoothing kernel
is a measure of diffusion and decay which  captures the time dependence mentioned
above.  
Other parameters (e.g. density, magnetic field strength, etc\ldots) can
affect the permeability of the ISM to CR electrons and impede their
propagation;
such parameters will vary with CR electron location, from the vicinity
of star-forming complexes to the more diffuse regions of the ISM.
As the image-smearing picture appears to provide the best quantitative
description for the FIR-radio correlation, we will now introduce
a more sophisticated approach to take into account the differences 
among CR electron populations within galaxy disks.

\section{Two-Component Image-Smearing Analysis \label{sec-multsmear}}
Using a phenomenological image-smearing model, first presented by
\citet{bh90}, M06b studied how the spatial distributions of FIR and
radio emission varied as a function of star-formation intensity within
12 spiral galaxy disks.
Their results were then related to the physical model of
\citet{hb93}.  
Although this method was found to improve the spatial correlation
between the distributions of infrared and radio emission by reducing
the residuals in the ratio maps,
a single kernel failed to work perfectly for an entire galaxy disk as
evidenced by the recognizable structures (i.e. spiral-arms) left in
the residual maps.  
This is not too surprising as the freshly injected CR electron
populations associated with star-forming regions should be
different in mean age and energy than those populations associated
with a galaxy's diffuse disk;  
such CR electrons have likely lost a significant amount of their
initial energy and/or have undergone re-acceleration by passing
interstellar shocks \citep[e.g.][]{sm98, egb03}. 
Consequently, it appears that a more realistic description should
require a multi-scale analysis of the galaxy images allowing for the
separation of structures at various spatial frequencies.   
We now present such an analysis using a wavelet-based
image-decomposition technique and compare these results to those using
a single-component method.

The simplest way to formulate a phenomenological model with two
distinct CR electron populations is to create representative source
functions and allow them to be characterized by different
scale-lengths.   
We therefore decompose the infrared images (used as source proxies)
into two components;
one representing sites of ongoing star formation and the other
consisting of diffuse emission from dust in the disk heated by older
stars or from afar by young stars.  
While a purely spatial definition will not yield a perfect census
of massive star formation sites, it has the advantage of simple
assumptions and data manipulation.

\subsection{Choice of Separation Scale}
The goal of the decomposition is to separate each 70~$\micron$ image
into two components: one containing the smallest to largest physical
structures associated with massive star formation and another to
include a galaxy's diffuse FIR emission distribution. 
The typical distance for which $>$90\% of Lyman continuum photons are
expected to be absorbed by interstellar gas and dust is $\sim$500~pc
\citep{ds94};   
this value is in excellent agreement with the observed mean-distances
of diffuse ionized gas surrounding H\textsc{II} regions \citep{af96}.
One kpc is also a moderate upper-limit to the sizes of giant
H\textsc{II} regions and molecular cloud complexes, as well as
widths of spiral arms.
Furthermore, the results presented by M06b hinted that the best-fit
global scale-lengths seemed to be spread around a median value of
$\sim$1~kpc, but did not cluster tightly around it.
For these reasons we choose a disk-structure separation scale of
1~kpc. 
While this scale may not be perfect for all galaxies, it is
physically motivated and allows for a uniform comparison among the
nearest 18 WSRT-SINGS sample galaxies.

\subsection{Wavelet Based Image Decomposition \label{sec-wt}}
The wavelet transform provides the ability to describe structures
within an image as a function of their characteristic spatial scale.
This property of the wavelet transform has led to the development of
the wavelet based Multi-scale Vision Model \citep[MVM;][]{br95,rb97}, a
procedure useful for identifying morphological features in
astronomical images \citep[e.g.][]{jls00,ca05}.
Wavelet analyses have also been used in to study the FIR-radio
correlation within individual systems such as NGC~6946 \citep{pf01},
the LMC \citet{ah06}, and M~33 \citep{ft07}; 
in each of these studies a cross-correlation analysis was performed on
the wavelet power spectra for images acquired at various wavelengths. 

\begin{figure*}
  \plotone{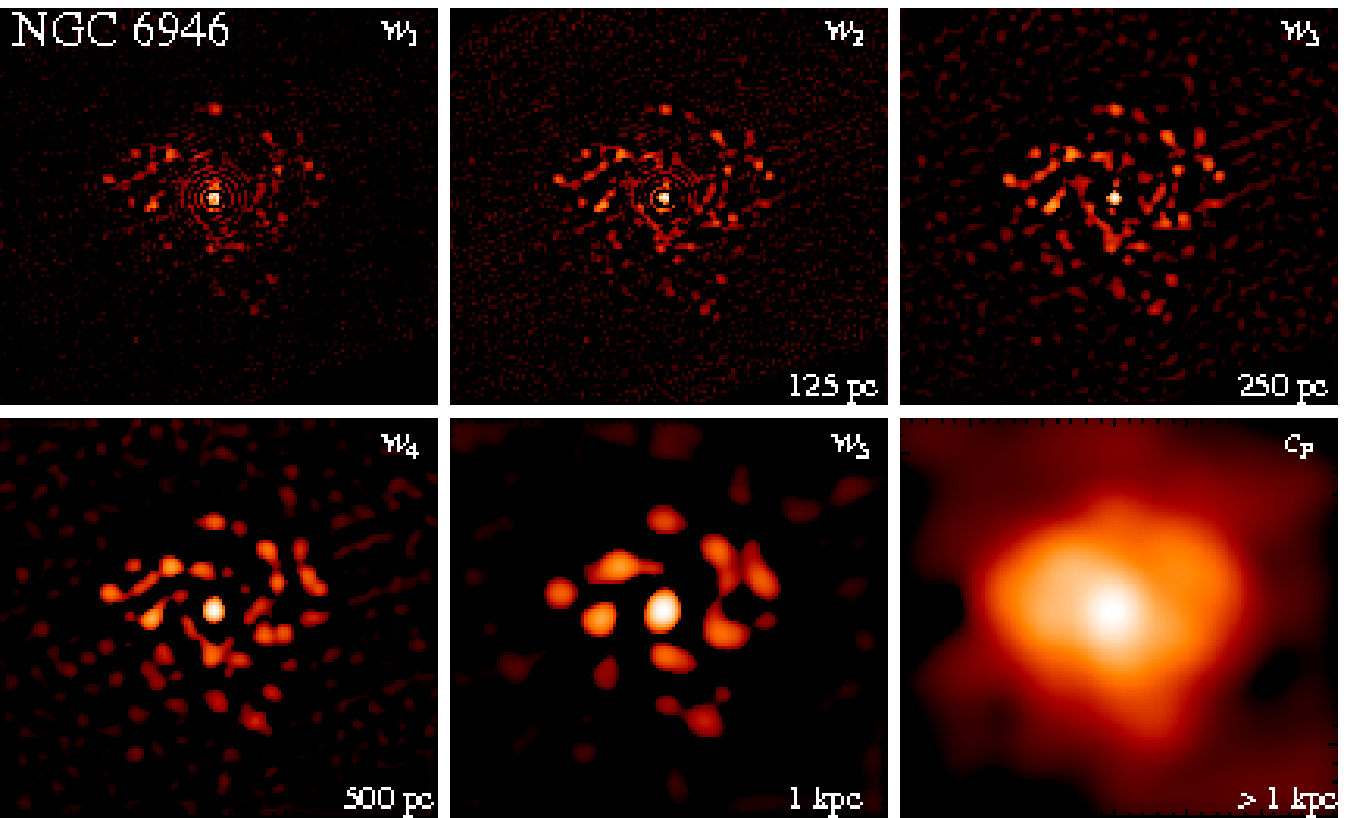}
  \caption{The wavelet transform of NGC~6946 at 70~$\micron$ using the
  {\it \`{a} trous} algorithm (see $\S$\ref{sec-wt}).  
  In this instance the image was re-gridded such that the each wavelet
  plane $w_{p=1-5}$ corresponds to 2$^{p-1}\times 62.5$~pc and the
  final smoothed plane, $c_{\rm P}$ contains information for the
  remaining larger spatial scales.   
  For the actual creation of the structure and disk images used in
  the present analysis, the re-gridding scheme described in
  $\S$\ref{sec-wt}  was used which ensures that the structure image is
  sampled at the Nyquist frequency. 
  Each plane is displayed using a logarithmic stretch ranging
  from the 1-$\sigma$ RMS level of the background to the maximum value
  of the wavelet coefficients at that plane.   
  The observed 70~$\micron$ image is recovered exactly by summing
  these six planes.
  \label{exwt}}
\end{figure*}

For the purpose of this study, we need only make use of the wavelet
transform to separate structures in each image as a function of their 
characteristic spatial frequency.
We apply the wavelet transform to our images using the 
{\it \`{a} trous} algorithm \citep{mhold89,ab91}. 
The {\it \`{a} trous} algorithm produces a set of wavelet coefficients
${w_{p}}$, at each scale $p$, having the same number of pixels as the
original image, $c_{o}$.
The exact physical scale corresponding to each wavelet plane, as 
indexed by $p$, will depend on the distance to the object and the pixel
scale of the image.  
It then follows that a pixel $j$ of the observed image can be
expressed as the sum all the wavelet coefficients at this position,
plus a final smooth plane containing the remaining large scale
features, $c_{\rm P}$, such that,
\begin{equation}
c_{o,j} = c_{{\rm P},j} + \sum_{p=1}^{\rm P} w_{p,j}
\end{equation}
An example of the wavelet transform via the {\it \`{a} trous}
algorithm is given in Figure \ref{exwt} for NGC~6946.
The complete {\it \`{a} trous} algorithm, as well as detailed
discussions on the applicability of wavelet transforms to astronomical
data sets, can be found in \citet{sm02}.

\begin{figure*}
  \plotone{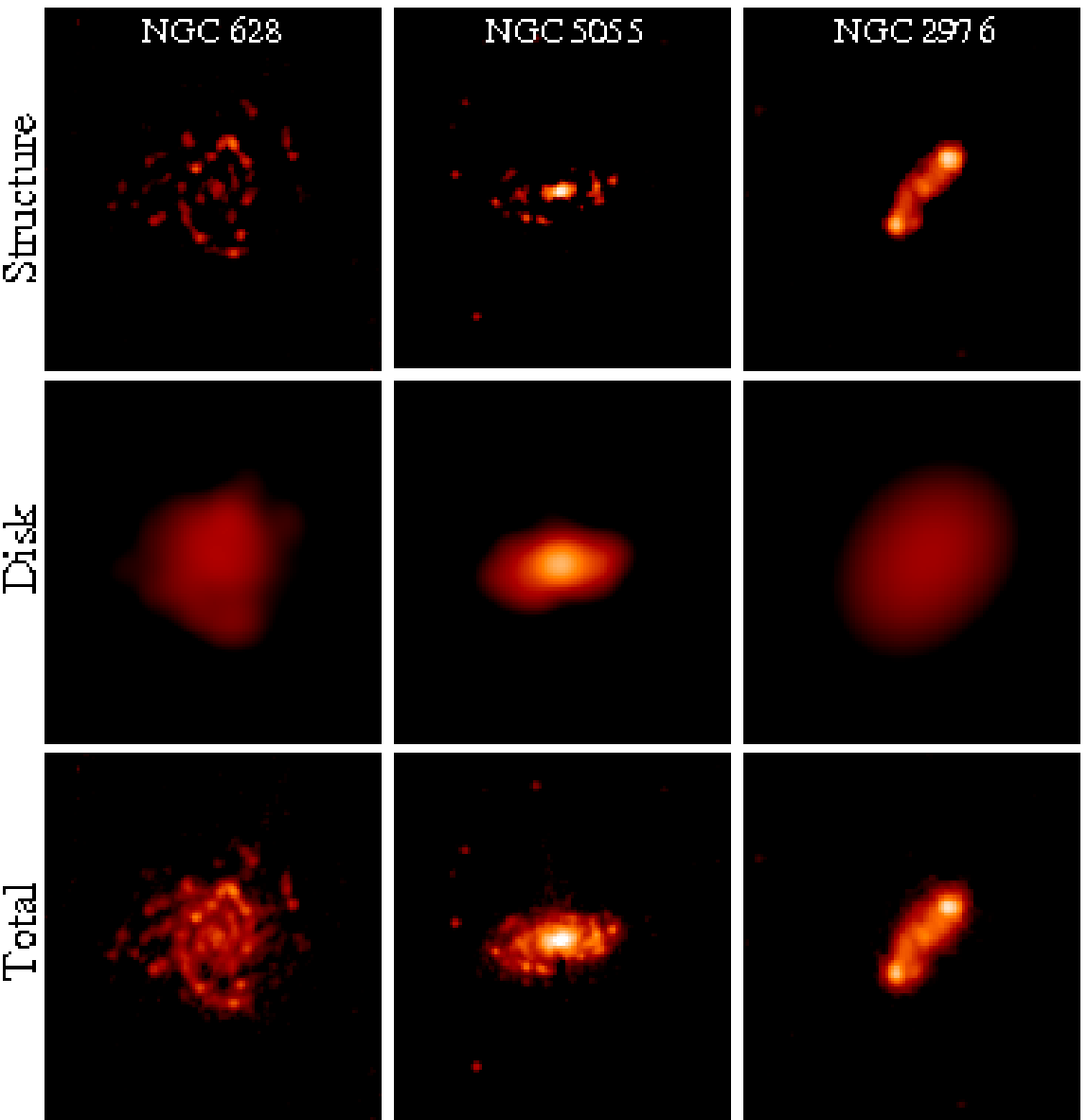}
  \caption{
    The {\it structure}, {\it disk}, and total (observed) 70~$\micron$
    images of NGC~628, NGC~5055, and NGC~2976.  
    These galaxies were chosen to span our sample's range in infrared
    surface brightness and are ordered, left to right, by increasing
    radiation field energy density.
    We have used a common logarithmic scaling for all six panels
    running from the 3-$\sigma$ level in the noisiest map to the
    maximum surface brightness level among all the maps. 
    The {\it structure} image contains all features significant at spatial
    frequencies less than 1~kpc while the {\it disk} image contains all
    features significant at spatial scales greater than or equal to
    1~kpc. (See $\S$\ref{sec-wt} for details.)
  \label{exstrdsk}}
\end{figure*}

We decompose each observed infrared image, $I$, into two sub-images:
(1) a {\it structure} image containing features with spatial 
scale smaller than 1~kpc; (2) a {\it disk} image containing all
structures with characteristic spatial scales larger than or equal to
1~kpc, largely constituting a galaxy's diffuse disk.  
The {\it \`{a} trous} algorithm is dyadic, transforming the image
at discrete scales which are powers of 2 and indexed here by $p$.   
We therefore re-grid each image appropriately such that 2$^{p-1}$
pixels, where \(p \geq 3\), corresponds to 1~kpc at a given galaxy's
distance.    
This re-gridding scheme ensures that the structure image is sampled at
the Nyquist frequency and provides uniformity in the decompositions for
all galaxies.  
The final structure and disk images, $I_{\rm str}$ and $I_{\rm dsk}$,
respectively, are then created by summing the appropriate scales such
that 
\begin{equation}
I_{\rm str} = \sum_{p=1}^{p(< 1~{\rm kpc})}w_{p}
\end{equation}
and
\begin{equation}
I_{\rm dsk} = I_{\rm P} + \sum_{p(\ge1~{\rm kpc)}}^{P}w_{p},
\end{equation}
where $I_{\rm P}$ is the final smooth plane containing all remaining
large scale features  and \(I = I_{\rm str} + I_{\rm dsk}\). 
For illustration, we display in Figure \ref{exstrdsk} the {\it
  structure} and {\it disk} images at 70~$\micron$ for three galaxies 
(NGC~628, NGC~5055, and NGC~2976) chosen to span a large fraction of
our sample's range in infrared surface brightness.
Each panel is displayed with the same stretch to allow comparison
between the morphology and amount of power residing in the disk and
structure components as a function of infrared surface brightness.
%

\subsection{Correcting for Free-Free Emission \label{sec-ffcor}}
The image-smearing analysis is designed to compare the
spatial distributions of a galaxy's FIR and non-thermal radio
emission. 
Because the observed radiation at 22~cm is a combination of
non-thermal (synchrotron) and thermal (free-free) emission, we
estimate and subtract the thermal fraction of radio emission from the
observed 22~cm maps.
This is done by estimating the free-free emission using a scaled
version of 24~$\micron$ maps as described in M06b.

The scaling factor relies on the empirical correlation found to exist
between 24~$\micron$ and extinction corrected Pa$\alpha$ luminosities
within NGC~5194 (M51a) by \citet{dc05}.
While it has been demonstrated that this correlation is not universal
\citep[e.g.][]{ppg06,dc07}, it has proven
to be more than sufficient as a first order estimate of the thermal
radio emission (M06b).    
Having estimated the Pa$\alpha$ line emission from the observed 24~$\micron$ emission, 
we can then determine the corresponding ionizing photon rate \citep{ost89} and expected free-free emission at 1.4~GHz \citep{rub68} such that, 
\begin{equation}
\left(\frac{S_{\rm T}}{\rm Jy}\right) \sim 7.93\times10^{-3}
\left(\frac{T}{10^{4}~{\rm K}}\right)^{0.45} 
\left(\frac{\nu}{\rm GHz}\right)^{-0.1}
\left(\frac{f_{\nu}(24~\micron)}{\rm Jy}\right),
\end{equation}
where we have assumed an average H~\textsc{II} region temperature of
\(T=10^{4}\)~K.

\begin{deluxetable}{ccc}
  \tablecaption{Radio Thermal Fraction Comparison \label{tfcomp}}
  \tablehead{
    \colhead{} & \colhead{$f_{\rm T}^{\rm 1~GHz}$} 
    & \colhead{$f_{\rm T}^{\rm 1~GHz}$}\\
    \colhead{Galaxy} & \colhead{\citet{nkw97}} & \colhead{This Paper}
  }
  \startdata
   NGC~3031  &0.05  &0.06\\
   NGC~3627  &0.09  &0.09\\
   NGC~4254  &0.05  &0.05\\
   NGC~4321  &0.05  &0.06\\
   NGC~4569  &0.07  &0.07\\
   NGC~4631  &0.03  &0.04\\
   NGC~4736  &0.18  &0.10\\
   NGC~5033  &0.11  &0.05\\
   NGC~5055  &0.09  &0.08\\
   NGC~5194  &0.05  &0.06\\
   NGC~6946  &0.06  &0.07\\
   NGC~7331  &0.06  &0.06
   
   \enddata
\end{deluxetable}

To check the reliability of this method for estimating the radio
thermal fractions, we compare our results to those of \citet{nkw97} for
12 galaxies which appear in both samples.
The thermal fractions of \citet{nkw97} were derived by the completely
independent method of radio spectral index fitting.
We list the 1.0~GHz thermal fractions of \citet{nkw97}, along with our
estimated thermal fractions, after scaling our 1.4~GHz flux densities
to what is expected at 1.0~GHz assuming a mean spectral index of
$-0.8$ for the non-thermal component, in Table \ref{tfcomp}.  
We find that, except for the cases of NGC~4736 and NGC~5033, our
estimated thermal fractions are very similar to those reported by
\citet{nkw97}.
We note that NGC~5033 hosts a Seyfert-2 type nucleus which could
complicate this method of free-free estimation using the 24~$\micron$
imaging.

\subsection{Phenomenological Image-Smearing Model \label{sec-smmod}}
The basic procedure used here is similar to that presented in M06a,b.
We calculate the residuals between the free-free corrected radio and
observed infrared images after convolving the infrared maps by a 
parameterized kernel  $\kappa({\bf r})$.
The new element is that we now smear $I_{\rm str}$ and $I_{\rm dsk}$
independently, then add the two smeared images and compare the sum to
the radio image.
The smoothing kernel is a function of a two-dimensional position
vector ${\bf r}$, having a magnitude \(r = (x^{2} + y^{2})^{1/2}\),
where $x$ and $y$ are the right ascension and declination offsets on
the sky. 

Due to the large range in inclination among our sample galaxies, we
compare results for exponential smoothing kernels oriented in either
the plane of the galaxy disk or isotropically.  
An exponential kernel was chosen because it was found to work as well
as, or better than, Gaussian kernels by M06a. 
This is consistent with ``leaky box'' type models where CR electron
escape is expected to occur on timescales less than or comparable to
the diffusion timescales \citep{bh90}.
The kernel takes the form,
\(\kappa({\bf r}) = e^{-{\bf r}/{\bf r_{\rm o}}}\),
where ${\bf r_{\rm o}}$ is its $e$-folding length $l$ modified by geometric factors related to the position angle $\theta$ and inclination $i$ of a galaxy such that 
\begin{equation}
{\bf r_{\rm o}} = \frac{l\cos i}{[1 - ({\bf x}\sin \theta +
   { \bf y}\cos\theta)^{2}\sin^{2}i/r^{2}]^{1/2}}. 
\end{equation}
The position angle $\theta$ of the tilt axis of the galactic disk is measured East of North and $i=0$ defines a face-on projection.  
Kernels are isotropic when $r_{\rm  o} = l$.  
This situation is the same as if a galaxy's inclination and position
angle were equal to zero.

\begin{figure*}[!ht]
\centerline{\hbox{
  \resizebox{18cm}{!}{
    \plotone{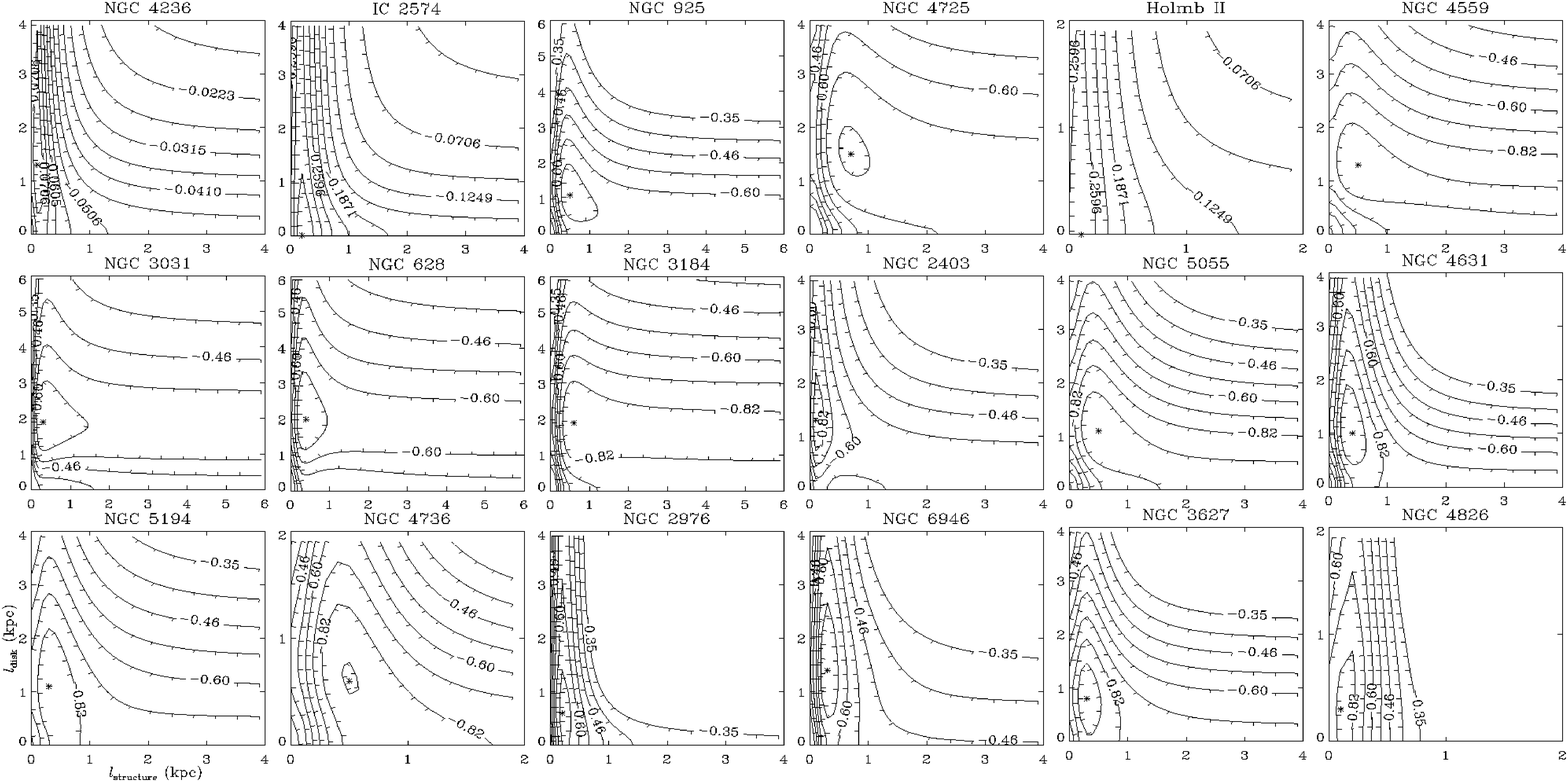}}}}
  \caption{
    The residual surfaces for each galaxy which are ordered
    by increasing $U_{\rm rad}$ from left to right, top to bottom.
    The abscissa and ordinate designate the scale-length of the kernel
    used to smooth the {\it structure} and {\it disk} component images,
    respectively.
    The minimum value on the surface is identified with an asterisk.
    A diagonal cut (i.e. where $l_{\rm str} = l_{\rm dsk}$) through
    each surface plot corresponds to the one dimensional residual
    curves which are plotted in figure \ref{rescurves}.
  \label{ressurfs}}
\end{figure*}

\begin{figure}[!ht]
    \plotone{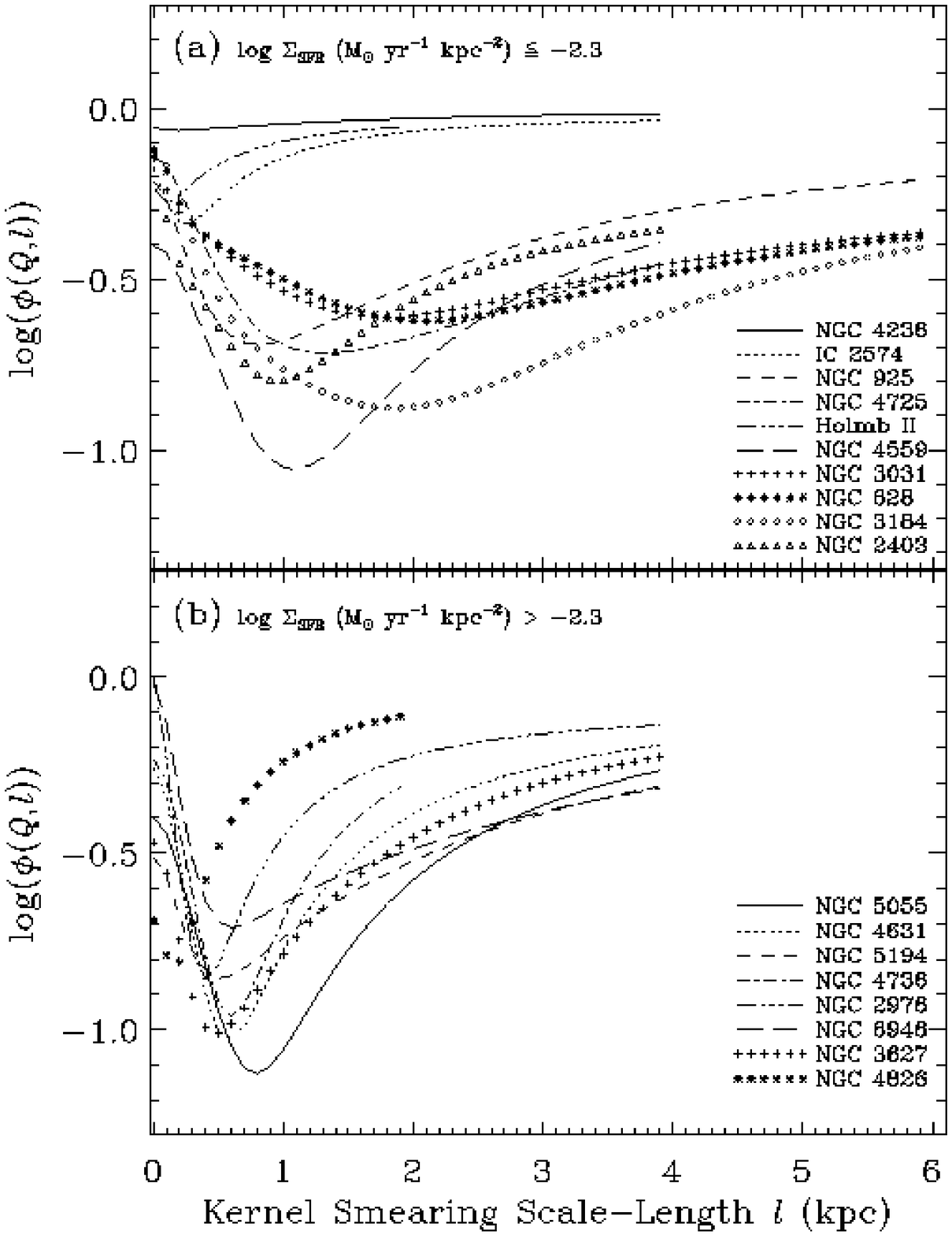}
  \caption{
  Residuals between the radio maps and smeared 70~$\micron$
  images (as defined in $\S$\ref{sec-smmod}) as a function of smearing
  scale-length.  
  The results plotted here use isotropic kernels 
  and 22~cm maps which have been corrected for the presence of
  free-free emission (see $\S$\ref{sec-ffcor}).
  Galaxies having low disk-averaged star formation rates, defined
  by \(\log~U_{\rm rad} \leq -12.5 = \log~\Sigma_{\rm SFR} \leq -2.3\)
  (see $\S$\ref{sec-sfrd}) are plotted in panel (a) while galaxies with
  high star formation activity, and larger values of $U_{\rm rad}$
  ($\Sigma_{\rm SFR}$), are plotted in panel (b).
  \label{rescurves}}
\end{figure}

Let $R({\bf r})$ denote the observed radio image and \(I({\bf r}) =
I_{\rm str}({\bf r}) + I_{\rm dsk}({\bf r})\) denote the observed
infrared image, which has been separated into component images
containing structures larger or smaller than $\sim$1~kpc (see
$\S$\ref{sec-wt}).  
The two-dimensional residual function, defined by $\phi$, is
calculated between the radio and smeared infrared images after the
infrared disk and structure images are first smoothed by independent
kernels and then summed.  
This residual surface is given by
\begin{equation}
\phi(Q,l_{\rm str},l_{\rm dsk}) =
\frac{\sum[Q^{-1}\tilde{I}_{j}(l_{\rm str},l_{\rm dsk}) - R_{j}]^{2}}
{\sum R_{j}^2},
\label{phi}
\end{equation}
where \(Q=\sum{I_{j}({\bf r})}/\sum R_{j}({\bf r})\) is
used as a normalization factor [i.e. \(\log~Q = q_{\lambda}(global)\)],
\begin{equation}
\tilde{I}(l_{\rm str},l_{\rm dsk}) = \tilde{I}_{\rm str}(l_{\rm str}) + 
\tilde{I}_{\rm dsk}(l_{\rm dsk}) 
\label{smIRimg}
\end{equation}
represents the infrared image after smearing $I_{\rm str}$ and $I_{\rm
  dsk}$ with independent kernels having scale length $l_{\rm str}$ and
$l_{\rm dsk}$, respectively, and the subscript $j$ indexes each
pixel.
Our calculation of the residuals is slightly different than what has
been used in M06a,b; we now fix the normalization factor, $Q$, rather
than let it vary with the kernel size.
This modification is necessary since we are now using different
smoothing kernels for each component.  
The quantity $\phi$ was minimized to determine the best-fit smearing
kernel for the structure and disk images of each galaxy in our
sample. 
The normalization by the squared sum of the radio flux density
allows for proper comparison of our galaxies which vary in intrinsic
surface brightness.  
We plot $\log~\phi$ as a function of $l_{\rm str}$ and $l_{\rm dsk}$
in Figure \ref{ressurfs}.
The special case of a single kernel, as studied by M06a,b, is
reproduced by the diagonal cut of 
of the residual surface plotted in Figure \ref{ressurfs} (i.e. when
\(l_{\rm  str} = l_{\rm dsk}\)).  
These residual curves are presented in Figure \ref{rescurves} and
binned by the amount of star formation activity within each galaxy.

The estimation of the residuals is carried out after first removing
pixels that are not detected at the 3-$\sigma$ level in either the
radio or maximally-smeared infrared maps, and editing out
contaminating background radio sources.
Additional editing was necessary for NGC~3031, NGC~3627, NGC~4725,
NGC~4736, and NGC~4826.  
Each of these galaxies host an AGN.  
Because the phenomenology we are trying to test is strictly
associated with the physics of star formation processes, we try to
remove any contamination an AGN might introduce.
This is done by first subtracting a scaled PSF from the nuclear region
of the 70~$\micron$ images before smoothing to remove any excess
emission associated with the AGN.
We then excise a 1~kpc region around the nucleus before calculating
the residuals between the smeared-infrared and radio images. 
Since the AGN within SINGS galaxies are all very low power, a 1~kpc
region should be adequate to remove the influence of an AGN.
Additional editing was also necessary for NGC~5194; its companion
galaxy (NGC~5195) was removed before calculating the residuals.

We use the quantity,
\begin{equation}
\Phi = \log~\left(\frac{\phi(Q,0,0)}{{\rm
    min}[\phi(Q,l_{\rm str},l_{\rm dsk})]}\right), 
\end{equation}
which is the depth of the minimum value in the residual surface,
as a measure of how much the correlation is improved by smoothing the
infrared structure and disk images.
In the special case when \(l_{\rm str} = l_{\rm dsk}\), we have 
\begin{equation}
\Phi_{\rm glob} = \log~\left(\frac{\phi(Q,0)}{{\rm
    min}[\phi(Q,l_{\rm str}=l_{\rm dsk})]}\right), 
\end{equation}
which is the same $\Phi$ studied by M06a,b (i.e. the minima found for
the diagonal residual curves presented in Figure \ref{rescurves}).   
We denote the best-fit scale-length which maximizes $\Phi_{\rm glob}$
as $l_{\rm glob}$ and refer to this quantity as the best-fit {\it
  global} scale-length for the galaxy.  
Uncertainties in $\phi$ are estimated by numerically propagating the
uncertainties in the input 70~$\micron$ and 22~cm images as measured
by the 1-$\sigma$ RMS noise of each map;
the uncertainty on the best-fit scale-length is then estimated as the
range in scale-length corresponding to the range from min($\phi$) to
min($\phi$)$+$unc($\phi$), spanning along the diagonal and structure
and disk axes of the residual surface maps.

While $\phi$ and $\Phi$ characterize the residual behavior as single
quantities, we also constructed residual images for each galaxy using
the best-fit smearing kernels to inspect the spatial variations of the
residuals using this new, two-component method.
The residual images, defined by,
\begin{equation}
{\rm Residual~image} = \log~(Q^{-1}\tilde{I}({\bf r})) - \log~(R({\bf r})),
\end{equation}
are plotted in column 5 of Figure \ref{resmaps} for isotropic kernels
having an exponential profile as this kernel type was associated with
the largest values of $\Phi$ (see $\S$\ref{sec-inc}).  
For a comparison, the residual maps associated with the best-fit
global scale-lengths are given in the fourth column of Figure
\ref{resmaps}.

\begin{figure}[!ht]
  \plotone{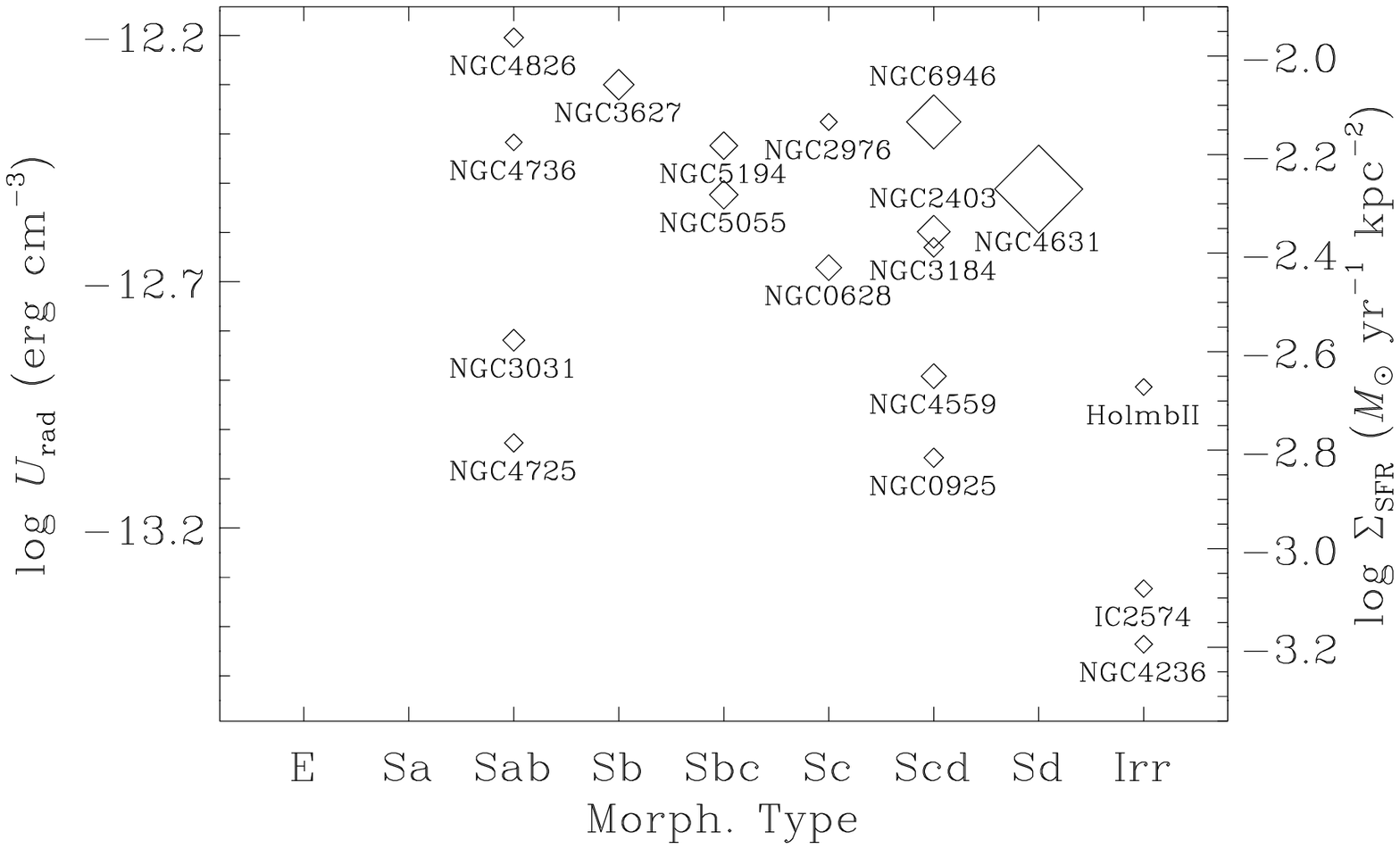}
  \caption{
    Radiation field energy density binned by morphological type.
    The plotting symbol sizes are scaled as a function of the
    improvement found using the two-component image-smearing procedure
    compared to using the single-component method (i.e. $\Phi
    -\Phi_{\rm glob}$).
    This value ranges from 0.00~dex for NGC~4736 to 0.24~dex for
    NGC~4631.  
   \label{dphimorph}}
\end{figure}

An additional free parameter of the relative amplitude of the disk and structure components was considered for a subset of the sample spanning the full range in surface brightness. 
This was done to test whether the proposed two-component analysis
alone is sufficient at capturing the signatures of CR electrons
associated with a galaxy's structures and disk or whether the relative
efficiencies between structure and disk processes in producing
synchrotron emission is significant.  
This could be especially important in the cases for which the power in
the structure and disk components are comparable. 
While slight quantitative improvements (i.e. on the few percent level)
are seen, they were not dramatic enough to warrant extra free
parameters.  

\section{Two-Component Modeling Results \label{sec-multresults}}
While we have shown that the FIR/radio ratios measured within galaxies
are most sensitive to star formation activity in $\S$4, we note that
the largest quantitative improvement to the dispersion in the FIR/radio
ratio within galaxies is achieved via our image-smearing analysis
(i.e. nearly a factor of 2 compared to a factor of $\sim$1.2 using a
residual dispersion analysis);
this procedure describes the spreading and decay of CR electrons in galaxy disks.
In fact, by repeating the correlation analysis described in $\S$4
with the best-fit smoothed infrared maps we find that the gradients
seen between $q_{70}$ and 24~$\micron$ surface brightness and
radius are flattened, on average, by factors of 5 and 4,
respectively.  
Hence, our phenomenological model clearly captures the behavior
observed in $\S$4;  
infrared emission is more strongly peaked around star-forming regions
since CR electrons diffuse away from bright structures into diffuse
regions of the ISM.  

In the following section we present the results from our two-component
image-smearing analysis and look for the cases in which significant
gains are obtained over the single-component method.
Since this phenomenology is valid for non-thermal radio continuum
emission, we focus our discussion on the results using radio images
that have been corrected for the presence of free-free emission as
described in $\S$\ref{sec-ffcor} and note that this correction only
slightly increases (i.e. 0.1~kpc on average) the measured best-fit
scale-lengths.


\begin{figure*}[!ht]
  \resizebox{18cm}{!}{
    \plotone{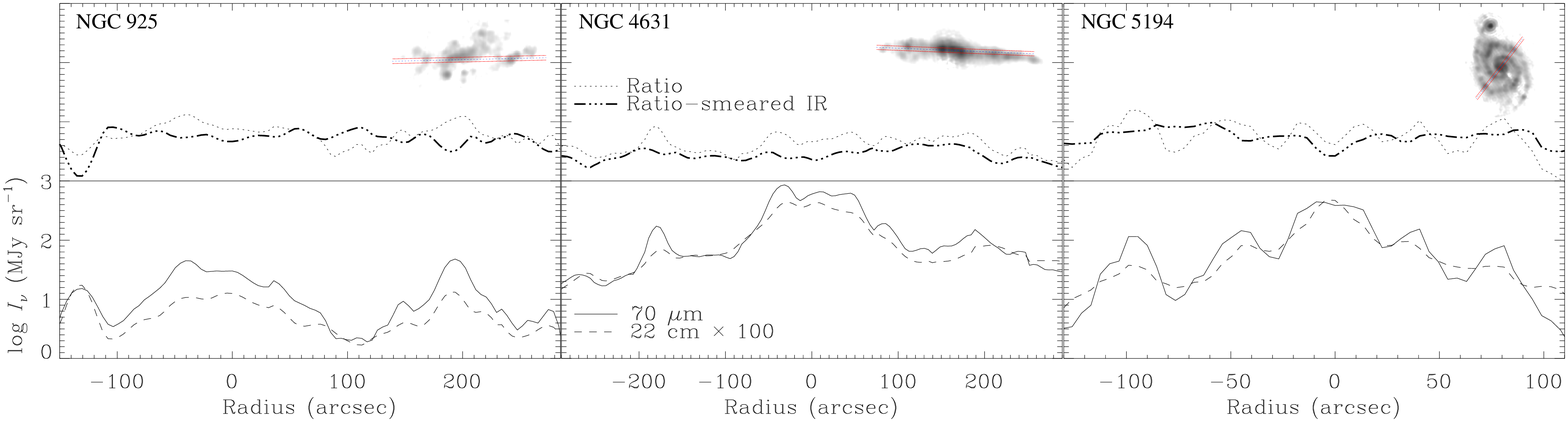}}
\caption{
  In the bottom portion of each panel we plot radial cuts across
  the 70~$\micron$ and 22~cm maps of NGC~925, NGC~4631, and NGC~5194.
  The 22~cm profile has been scaled by a factor of 100 to allow for a
  better comparison against the 70~$\micron$ profile.
  In the top portion of each panel we plot the ratio between the
  radio and the observed infrared profiles along with the ratio between the radio the infrared profiles after smoothing by the best-fit kernels.
  Each cut, displayed in the thumbnail image of the galaxy, has a
  width equal to the FWHM of the 70~$\micron$ PSF.} 
\label{slice}
\end{figure*}

\subsection{Comparison with Single-Component Method \label{sec-mscomp}}  
We find a range in the behavior of the residual surface maps among the
sample galaxies in Figure \ref{ressurfs}.
For nearly all galaxies we observe an initially steep (negative)
gradient for increasing structure scale-lengths as the minimum in
$\log~\phi$ is reached.    
Even though some galaxies may not exhibit a well defined minimum along
the structure axis (e.g. NGC~628), the steep initial gradient clearly
demonstrates that the structure component is being detected by our
two-component analysis. 
Similarly, many galaxies show a well defined minimum along the disk
axis (e.g. NGC~4559) while others do not (e.g. NGC~4631) suggesting
that this component is evident as well.  
This diverse behavior among the residual maps demonstrates that our
two-component analysis does yield qualitative insight into the two
major CR electron populations under discussion.

To assess quantitatively the gains of using the two-component approach
over image-smearing with a single kernel, we look at the differences
between $\Phi$ and $\Phi_{\rm glob}$ which we give in Table
\ref{scltbl}.  
We find a large range in improvement spanning from 0.00 (NGC~4736) to
0.24~dex (NGC~4631). 
While galaxies showing the largest improvements generally have high
infrared surface brightnesses, galaxies exhibiting very little
improvement are well distributed as a function of infrared surface
brightness. 
We also assess whether improvement correlates with morphology.

In Figure \ref{dphimorph} we plot each galaxy's radiation field energy
density as a function morphological type; 
the plotting symbol sizes are proportional to $\Phi -\Phi_{\rm glob}$.
Irregular and early-type spirals show essentially no improvement
by using the two-component approach.
Significant improvement appears limited to spiral galaxies of type Sb and
later with high surface brightnesses.

The negligible improvement found 
for irregulars ($\Phi - \Phi_{\rm glob}\la$0.005~dex) can be explained
by their lack of a diffuse emission component at both FIR and radio
wavelengths.
This is clearly seen by inspecting the 70~$\micron$ and 22~cm images
of these galaxies in Figure \ref{resmaps}.
As a consequence of this, these galaxies have been excluded in the
statistical comparisons of best-fit scale-lengths presented in
$\S$6.3 and $\S$6.5; these galaxies will be revisited in
$\S$\ref{sec-discevo} of the discussion.

As for the early-type spirals, small ($\Phi - \Phi_{\rm glob}
\la$0.02~dex) improvements are probably expected since their FIR
and radio emission are dominated by an inner-disk;  
if the central concentration of emission largely dominates over that
of the disk, this morphology will be pretty well described by a
single-component compared to a galaxy disk containing many organized
star-forming structures.  
Consequently, we find that the two-component approach has mainly
improved our phenomenological description of the FIR-radio correlation
for late-type galaxies exhibiting clear spiral structure and
significant star-formation throughout their disks; 
this corresponds to roughly half the sample spirals, for which $<\Phi -
\Phi_{\rm glob}> \sim$0.1~dex.

A discrepancy is found for NGC~2976, which is a member of the M~81
group.
This discrepancy may be due to the fact that the galaxy's optical
morphology (which is classified as an Sc peculiar) is strikingly
different than its morphology in the FIR.   
In the FIR its disk emission arises from a very small disk ($D_{\rm
  TIR} \sim$6~kpc) and is dominated by two large star-forming complexes
on either side of a relatively weak nucleus;
these features are almost unnoticeable in the optical.  
In fact, detailed studies of NGC~2976 carried out in the optical
\citep{bnm92} and in HI \citep{si02} have shown that the galaxy has
dwarf like characteristics and possibly triggered star formation due
to encounters between other members in the central M81 group.  
Furthermore, the optical luminosity, and presumably mass, of NGC~2976 is much smaller than that for the sample spirals, but similar to the sample irregulars.
Such findings are consistent with our result; 
NGC~2976 behaves more like the irregulars in the sample for which we
find little improvement is obtained by using the two-component analysis
compared to the single-component method.

\subsection{Morphological Differences in Residual Images
  \label{sec-resmaps}} 
The residual images using the single-component and two-component
analyses are displayed in the fourth and fifth columns of Figure
\ref{resmaps}, respectively.
As already stated in $\S$\ref{sec-mscomp}, we find that the
decrease in the residuals between the FIR and radio maps for the two-component analysis with respect to the single-component analysis is significantly greater for late-type spirals having a relatively large
amount of ongoing star formation.  

The residual images using the single-component method display a
general trend such that galaxies with low star formation activity
exhibit radio excesses associated with star-forming regions
while galaxies with higher star formation activity behave in an
opposite manner; 
these galaxies display infrared excesses associated with
star-forming regions in their residual images.  
This is the result of galaxies with low star-formation activity
generally needing large scale-lengths to best match their infrared and
radio spatial distributions leading to over-smoothed star-forming regions relative to the more actively star-forming galaxies.  
Now, by using the two-component technique and smoothing small spatial
scale features (i.e. individual star-forming complexes) with smoothing
functions independent of those used for the large spatial frequency
components, this trend has been largely been suppressed for a number
of galaxies (i.e. NGC~4631, NGC~5194, NGC~6946).

A comparison of the residual images for our sample's nearest
grand-design spiral (NGC~6946) in Figure \ref{resmaps}, best
illustrates this result; 
the rather dramatic differences in the residual behavior for
star-forming spiral arms and quiescent inter-arm regions, which are
evident when using a single kernel, are largely suppressed by our new
treatment. 
Some organized structure still remains in our two-component residual
image which is not surprising since this method only looks to
characterize the extremes of a galaxy's CR electron population.
The remaining structures might also suggest that our symmetric
smoothing functions have limited applicability;
diffusion preferentially occurring along field lines, which trace arms and become
tangled in star-forming regions, will lead to asymmetries associated
with spiral arm structure \citep[e.g. NGC~6946;][]{beck07}.  

Since it is difficult to determine amplitude information from the
residual images in Figure \ref{resmaps}, we show radial cuts across
the 70~$\micron$ and 22~cm maps, along with the ratios between the radio
and both the observed and smoothed 70~$\micron$ images, in Figure
\ref{slice}.
This is done for three galaxies (NGC~925, NGC~4631, and NGC~5194)
chosen to span our sample's morphology and inclination.  
These plots clearly illustrate that the amplitude of the variations in
the FIR/radio ratios across each galaxy disk is significantly reduced
by our image-smearing procedure.  
In the case of the grand design spiral NGC~5194, we also find that the
arms are broader in the radio compared to the FIR;
this observation is consistent with CR electrons being accelerated in
star formation sites within spiral arms and diffusing larger distances
than associated dust heating photons.

\begin{deluxetable}{ccccc|ccc}
  \tablecaption{Effects of Inclination on Kernel Shape for Sample
    Spirals\label{tbl-inc}} 
\tablehead{
  \multicolumn{1}{c}{Inclination}&
  \multicolumn{1}{c}{}&
  \multicolumn{2}{c}{median $\Phi$ (dex)}&
  \multicolumn{1}{c}{}&
  \multicolumn{2}{c}{median $\Phi_{\rm glob}$ (dex)}&
  \multicolumn{1}{c}{}\\
  \colhead{(deg)} & \colhead{$N$} &
  \colhead{iso} & \colhead{gal} & \colhead{$\delta \Phi$} &
  \colhead{iso} & \colhead{gal} & \colhead{$\delta \Phi_{\rm glob}$}
}
\startdata
$\leq 60\degr$ &11 &0.59  &0.57  &0.02    &0.57  &0.55  &0.02\\
$> 60\degr$    & 4 &0.87  &0.61  &0.17    &0.77  &0.56  &0.13
\enddata
\end{deluxetable}

\subsection{Inclination Effects \label{sec-inc}}
We now determine how inclination affects the choice of smoothing
kernel. 
In Table \ref{tbl-inc} we give the median values of $\Phi$ and
$\Phi_{\rm glob}$ for isotropic 
or in-disk kernels after separating the sample into low ($i \leq 60\degr$)
and high ($i > 60\degr$) inclination bins.  
As expected, the difference for face-on galaxies appears almost negligible;
the median \(\delta\Phi = {\rm med}(\Phi_{\rm iso}
- \Phi_{\rm gal})\) for these galaxies is $\sim$0.02~dex while
$\delta\Phi_{\rm glob}$ is also $\sim$0.02~dex. 
Conversely, if a galaxy's inclination is greater than $\sim$60$\degr$
then the orientation of the kernel seems to become important.
In this case, isotropic kernels are strongly favored such that $\delta\Phi$ is $\sim$0.17~dex while $\delta\Phi_{\rm glob}$ is $\sim$0.13~dex. 

The regular component of a galaxy's magnetic field is spread most
densely throughout its {\it thin} disk;
the radial diffusion of CR electrons should then preferentially occur
along field lines while vertical (out-of-plane) diffusion should
require CR electrons to undergo an increased amount of cross-field
diffusion.
Such a scenario has been verified empirically.
Indirect estimates of diffusion coefficients for the vertical
propagation of CR electrons in a galaxy's thin disk have been found to
be an order of magnitude smaller than those for radial diffusion
\citep{dlg95}.

However, in galaxies with active star formation, ordered magnetic
fields can be ruptured allowing CR electrons to quickly escape their
disks and form synchrotron haloes \citep[e.g.][]{eh88}. 
This decrease in vertical confinement will lead to a diffusion behavior
which is more isotropic in appearance.
A clear example of this scenario is seen in the nearly edge-on galaxy,
NGC~4631.
NGC~4631 has a large radio halo that extends $\sim$5~kpc beyond the
vertical extent of the FIR disk \citep{dlg95}.
Our results therefore suggest that the highly inclined sample galaxies
each possess, at least to some degree, synchrotron haloes in which the
diffusion of CR electrons occurs on similar time-scales as those in the disk.

\subsection{Best-fit Scale-lengths \label{sec-bfscl}}
In Figure \ref{ressurfs} we plot the residual surfaces associated with
the two-component image-smearing procedure as described in
$\S$\ref{sec-smmod}.  
The best-fit disk, structure, and global scale-lengths, along with
estimated errors (see $\S$\ref{sec-smmod}), are given in Table
  \ref{scltbl}. 
We find that the best-fit scale-lengths for the structure images are
on average $\sim$ 0.5~kpc while the best-fit scale-lengths for the disk
images are, on average, $\sim$1.3~kpc; or, in other words, the
best-fit disk scale-length is, on average, $\sim$2.5 times larger than
the best-fit structure scale-length.
The best-fit global scale-lengths, using the single-component
approach, are found to be $\sim$1.0~kpc, on average.

To ensure that the exact separation scale used in decomposition of the
70~$\micron$ images does not affect the best-fit disk and structure
scale-lengths we repeated our analysis using separation scales ranging
from 0.5 (for galaxies resolved at this scale) to 2.0~kpc for 4
galaxies of varying size, morphology, and infrared surface brightness
(i.e. NGC~628, NGC~2976, NGC~2403, NGC~6946).
We find that the dispersion in best-fit disk and structure
scale-lengths is $\sim$0.15 and 0.10~kpc, respectively.
Our separation method therefore seems quite robust and the best-fit
disk and structure scale-lengths appear to be fairly insensitive to
the exact separation scale. 

In Figure \ref{SFRDles-gd}a we plot the best-fit disk and structure
scale-lengths versus radiation field energy density, which we also
express as a star formation surface density.  
Also included in this plot are the best-fit scale-lengths found using
the single-component method.
We perform ordinary least squares fits to the best-fit disk, structure, and
global scale-lengths with $U_{\rm rad}$ as the independent
variable.  
The least square fits for the best-fit disk and structure
scale-lengths in Figure \ref{SFRDles-gd}a have very similar slopes
(-0.43 and -0.47, respectively).  
Furthermore, the best-fit global scale-lengths appear to be nearer to
the best-fit disk scale-lengths for galaxies having low values of
$U_{\rm rad}$ and nearer to the best-fit structure scale-lengths
for galaxies having high values of $U_{\rm rad}$.  
This result is consistent with another observation, namely that
the best-fit global scale-lengths display a general trend of
decreasing with increasing ratio between the power in the
70~$\micron$ structure to disk components.  
These results suggest that the difference between the FIR and
radio emission spatial distributions is dependent on whether
the disk or structure is the dominant emission component of a galaxy.  
In $\S$7 we propose that the disk will generally be composed of a population of 
CR electrons which are significantly older than those which have been 
recently injected into the ISM and still reside close to their parent
star-forming regions. 

We also note that four galaxies (NGC~925, NGC~4725, NGC~4559, and
NGC~3031) appear to be off of the general trend between best-fit
global scale-length and star formation rate surface density first
identified by M06b.  
While the best-fit structure scale-lengths of these galaxies appear
consistent with the rest of the sample, the best-fit disk and global
scale-lengths appear to be lower than what one would expect
compared to the trend for the higher surface brightness galaxies.
Of the 15 galaxies we plot in Figure \ref{SFRDles-gd}a, these four
galaxies also have the four lowest surface brightnesses.
This turnover at $\log U_{\rm rad} \sim -12.7$ may be due to a signal-to-noise effect;  
specifically, we may not be detecting the most diffuse emission in the
extended parts of these galaxy disks which is throwing off our
estimations for the best-fit scale-lengths.
For example, by truncating the areas over which we calculate the
residuals, the turnover is shifted towards values of $\log U_{\rm rad}
> -12.7$.
Conversely, simulations of much deeper observations to detect a very low
surface brightness disk lead to longer scale lengths, suggesting again
that the turnover may be due to the limited sensitivity of our data.
On the other hand, the location of these galaxies in Figure
\ref{SFRDles-gd}a may not be a signal-to-noise artifact, but due to
a different physical make-up of these systems.   
Because it is difficult to settle this issue without deeper
observations, we include a discussion on the physical implications of
this scenario in $\S$\ref{sec-discevo}.

\begin{deluxetable}{ccccccc}
  \tablecaption{Two-Component Fitting Results\label{scltbl}}
  \tablewidth{0pt}
  \tablehead{
    \colhead{} &
    \colhead{$l_{\rm glob}$}  &
    \colhead{$l_{\rm str}$}  &
    \colhead{$l_{\rm dsk}$} &
    \colhead{$\Phi_{\rm glob}$} & \colhead{$\Phi$} &
    \colhead{$\Phi - \Phi_{\rm glob}$}\\
    \colhead{Galaxy} &
    \colhead{(kpc)} & \colhead{(kpc)} & \colhead{(kpc)} &
    \colhead{(dex)} & \colhead{(dex)} & \colhead{(dex)}
    }
  \startdata
 NGC~0628   &2.1$^{+0.73}_{-0.61}$ &  0.5$^{+0.28}_{-0.18}$ &  2.3$^{+0.60}_{-0.52}$ &  0.499  &0.561  &0.063\\
 NGC~0925   &0.8$^{+0.33}_{-0.21}$ &  0.6$^{+0.37}_{-0.20}$ &  1.1$^{+0.57}_{-0.42}$ &  0.479  &0.501  &0.022\\
 NGC~2403   &1.0$^{+0.16}_{-0.21}$ &  0.3$^{+0.16}_{-0.10}$ &  1.4$^{+0.37}_{-0.32}$ &  0.664  &0.762  &0.098\\
 Holmb~II   &0.1$^{+0.10}_{-0.04}$ &  0.1$^{+0.10}_{-0.04}$ &  0.0$^{+1.44}_{\rm- NaN}$ &  0.043  &0.043  &0.000\\
 NGC~2976   &0.4$^{+0.01}_{-0.01}$ &  0.4$^{+0.03}_{-0.03}$ &  0.5$^{+0.16}_{-0.20}$ &  0.864  &0.867  &0.003\\
 NGC~3031   &1.8$^{+0.73}_{-0.58}$ &  0.5$^{+0.59}_{-0.22}$ &  1.9$^{+0.66}_{-0.44}$ &  0.464  &0.503  &0.039\\
 NGC~3184   &1.9$^{+0.39}_{-0.45}$ &  0.9$^{+0.67}_{-0.26}$ &  2.0$^{+0.37}_{-0.35}$ &  0.741  &0.764  &0.023\\
 IC~2574    &0.2$^{+0.17}_{-0.03}$ &  0.3$^{+0.11}_{-0.13}$ &  0.0$^{+0.60}_{\rm- NaN}$ &  0.161  &0.167  &0.006\\
 NGC~3627   &0.5$^{+0.01}_{-0.01}$ &  0.4$^{+0.01}_{-0.03}$ &  0.8$^{+0.13}_{-0.07}$ &  0.541  &0.629  &0.088\\
 NGC~4236   &0.2$^{+0.26}_{-0.16}$ &  0.1$^{+0.12}_{-0.12}$ &  1.4$^{+3.11}_{-1.02}$ &  0.009  &0.017  &0.008\\
 NGC~4559   &1.1$^{+0.03}_{-0.05}$ &  0.6$^{+0.06}_{-0.11}$ &  1.4$^{+0.06}_{-0.14}$ &  0.655  &0.713  &0.058\\
 NGC~4631   &0.6$^{+0.05}_{-0.09}$ &  0.4$^{+0.01}_{-0.00}$ &  1.1$^{+0.05}_{-0.06}$ &  0.768  &1.009  &0.241\\
 NGC~4725   &1.3$^{+0.84}_{-0.47}$ &  0.9$^{+3.94}_{-0.41}$ &  1.6$^{+0.98}_{-0.71}$ &  0.570  &0.585  &0.015\\
 NGC~4736   &0.6$^{+0.02}_{-0.01}$ &  0.6$^{+0.03}_{-0.04}$ &  0.6$^{+0.11}_{-0.07}$ &  0.729  &0.729  &0.000\\
 NGC~4826   &0.2$^{+0.01}_{-0.04}$ &  0.2$^{+0.01}_{-0.04}$ &  0.5$^{+0.23}_{-0.25}$ &  0.117  &0.138  &0.021\\
 NGC~5055   &0.8$^{+0.06}_{-0.09}$ &  0.5$^{+0.01}_{-0.00}$ &  1.1$^{+0.02}_{-0.03}$ &  0.727  &0.805  &0.078\\
 NGC~5194   &0.5$^{+0.10}_{-0.04}$ &  0.4$^{+0.01}_{-0.05}$ &  1.1$^{+0.13}_{-0.14}$ &  0.343  &0.417  &0.073\\
 NGC~6946   &0.6$^{+0.14}_{-0.10}$ &  0.4$^{+0.02}_{-0.00}$ &  1.6$^{+0.10}_{-0.11}$ &  0.690  &0.860  &0.171\\
 \hline \hline
{\bf Averages}$^{a}$ &{\bf 0.9} &{\bf 0.5} &{\bf 1.3} &{\bf
  0.590}  &{\bf 0.656} &{\bf 0.066}
  \enddata
\tablecomments{$^{a}$: The irregular galaxies (Holmb~II, IC~2574, and
  NGC~4236) were excluded when calculating the averages. }
\end{deluxetable}

\section{Discussion \label{sec-disc}}
We find that the FIR and non-thermal radio morphologies are more
similar to each other for galaxies having higher radiation field
energy densities compared to galaxies with lower radiation field
energy densities.  
Following the interpretation of M06b, our results indicate that CR
electrons are, on average, closer to their place of origin in galaxies
having higher star formation activity.

Since the diffusion scale-length of CR electrons depends only on their
age and ability to diffuse through the ISM of galaxies, there are
four possible explanations of the results.  
The CR electrons within galaxies having high star formation activity
may: 
(1) have relatively short lifetimes due to a high energy loss
rate;
(2) diffuse at a slower rate due to the ISM having a high density
and magnetic field strength, resulting in a shorter mean free path;
(3) escape into intergalactic space at a higher rate due to an
increased occurrence of ruptured magnetic field lines (this
explanation implies systematically higher global FIR/radio ratios); or 
(4) have been accelerated recently and be relatively young.
The first three of these explanations are applicable in the case of
steady-state star formation (i.e. the variation of ISM parameters
alone will lead to shorter scale-lengths).
The fourth explanation, however, requires a recent episode of enhanced
star formation to inject fresh CR electrons into a galaxy's ISM. 
We will try to distinguish among these four scenarios in the
context of the results for the two-component smearing analysis.

\subsection{CR-Electron Cooling Timescales and Diffusion Properties in Normal Galaxies
\label{sec-discCR}} 
We now describe the physical processes associated with the propagation and decay of
CR electrons in the ISM.
We derive CR electron cooling timescales and introduce a diffusion equation to derive propagation distances and physical scaling relations for comparison with our observations. 
These comparisons will allow us to discriminate between the four scenarios above.

As CR electrons propagate through the ISM of galaxies they lose their
energy due to a number physical processes  
including synchrotron, inverse-Compton (IC) scattering,
bremsstrahlung, ionization, and adiabatic expansion losses.
In normal galaxies 
synchrotron and IC scattering processes are likely the most significant energy loss terms
for CR electrons associated with 1~GHz emission \citep{jc92}; 
the other terms listed will become non-negligible, however, for
galaxies hosting extreme episodes of star-formation like starbursting
ultra-luminous infrared galaxies (ULIRGs) \citep[e.g.][]{tt06}.  
We now derive the CR electron cooling timescales associated with the two dominant processes.  

Let us assume that CR electrons propagating 
with a pitch angle $\alpha$ in a magnetic field of strength $B$ have
isotropically distributed velocities such that \(<\sin^{2}\alpha> =
\twothirds\) leading to \(B_{\perp} \approx 0.82~B\).
According to synchrotron theory, a CR electron having energy $E$ will
emit most of its energy at a critical frequency $\nu_{\rm c}$ where  
\begin{equation} 
\label{eq-nuBE}
 \left(\frac{\nu_{\rm c}}{\rm GHz}\right) = 1.3\times10^{-2}
  \left(\frac{B}{\rm \mu G}\right)
  \left(\frac{E}{\rm GeV}\right)^{2}.
\end{equation}
Since the energy loss of CR electrons by synchrotron radiation goes as 
$dE/dt \propto U_{\rm B}E^{2}$, we can use Equation \ref{eq-nuBE} to
express the synchrotron cooling timescale, \(\tau_{\rm syn} \equiv
E/|dE/dt|_{\rm syn}\), for CR electrons as  
\begin{equation}
\label{eq-tsync_ub}
  \left(\frac{\tau_{\rm syn}}{\rm yr}\right) \sim 5.7\times10^{7}
  \left(\frac{\nu_{\rm c}}{\rm GHz}\right) ^{-1/2}
  \left(\frac{B}{\rm \mu G}\right)^{1/2}
  \left(\frac{U_{\rm B}}{10^{-12}~{\rm erg~cm^{-3}}}\right)^{-1},
\end{equation}
where \(U_{\rm B} = B^{2}/(8\pi)\) is the magnetic field energy
density.  
Naturally, we can rewrite the synchrotron cooling timescale as
\begin{equation}
  \label{eq-tsync}
  \left(\frac{\tau_{\rm syn}}{\rm yr}\right) \sim 1.4\times10^{9}
  \left(\frac{\nu_{\rm c}}{\rm GHz}\right) ^{-1/2}
  \left(\frac{B}{\rm \mu G}\right)^{-3/2}.
\end{equation}
Similarly, since the energy loss of CR electrons due to IC scattering
goes as $dE/dt \propto U_{\rm rad}E^{2}$, we can again use Equation \ref{eq-nuBE} to write the IC cooling
timescale, \(\tau_{\rm IC} \equiv E/|dE/dt|_{\rm IC}\), as
\begin{equation}
\label{eq-tIC}
  \left(\frac{\tau_{\rm IC}}{\rm yr}\right) \sim 5.7\times10^{7}
  \left(\frac{\nu_{\rm c}}{\rm GHz}\right) ^{-1/2}
  \left(\frac{B}{\rm \mu G}\right)^{1/2}
  \left(\frac{U_{\rm rad}}{10^{-12}~{\rm erg~cm^{-3}}}\right)^{-1}, 
\end{equation}
where  $U_{\rm rad}$ is the radiation field energy density.
Photons with a frequency $\nu_{\rm p}$ will significantly
contribute to IC losses if below the Klein-Nishina limit; 
for GeV electrons considered here, the upper photon energy limit lies
in the X-ray regime (i.e. $h\nu_{\rm p} \la 200~{\rm eV}$), but the
bulk of losses arise from interactions with IR/optical photons,
which dominate $U_{\rm rad}$.
The effective cooling timescale for CR electrons due to synchrotron
and IC losses is 
\begin{equation}
\label{eq-cool_shrt}
\tau_{\rm cool}^{-1} = \tau_{\rm syn}^{-1} + \tau_{\rm IC}^{-1},
\end{equation}
which, by combining Equations \ref{eq-tsync_ub} and \ref{eq-tIC},  we
can express as
\begin{equation}
\label{eq-cool}
  \left(\frac{\tau_{\rm cool}}{\rm yr}\right) \sim 5.7\times10^{7}
  \left(\frac{\nu_{\rm c}}{\rm GHz}\right) ^{-1/2}
  \left(\frac{B}{\rm \mu G}\right)^{1/2}
  \left(\frac{U_{\rm B}+U_{\rm rad}}{10^{-12}~{\rm
  erg~cm^{-3}}}\right)^{-1}.
\end{equation}

\begin{figure*}[!ht]
  \plottwo{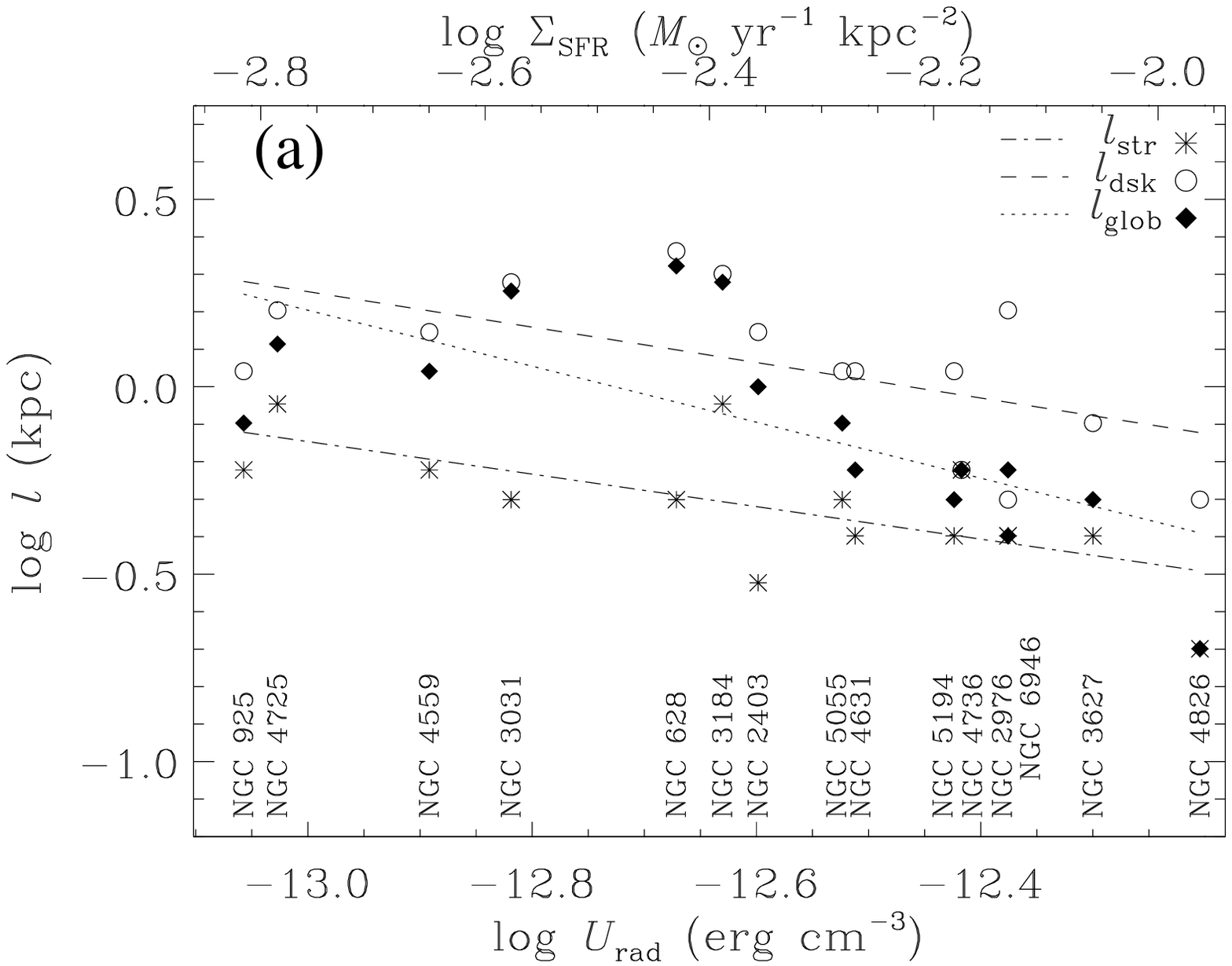}{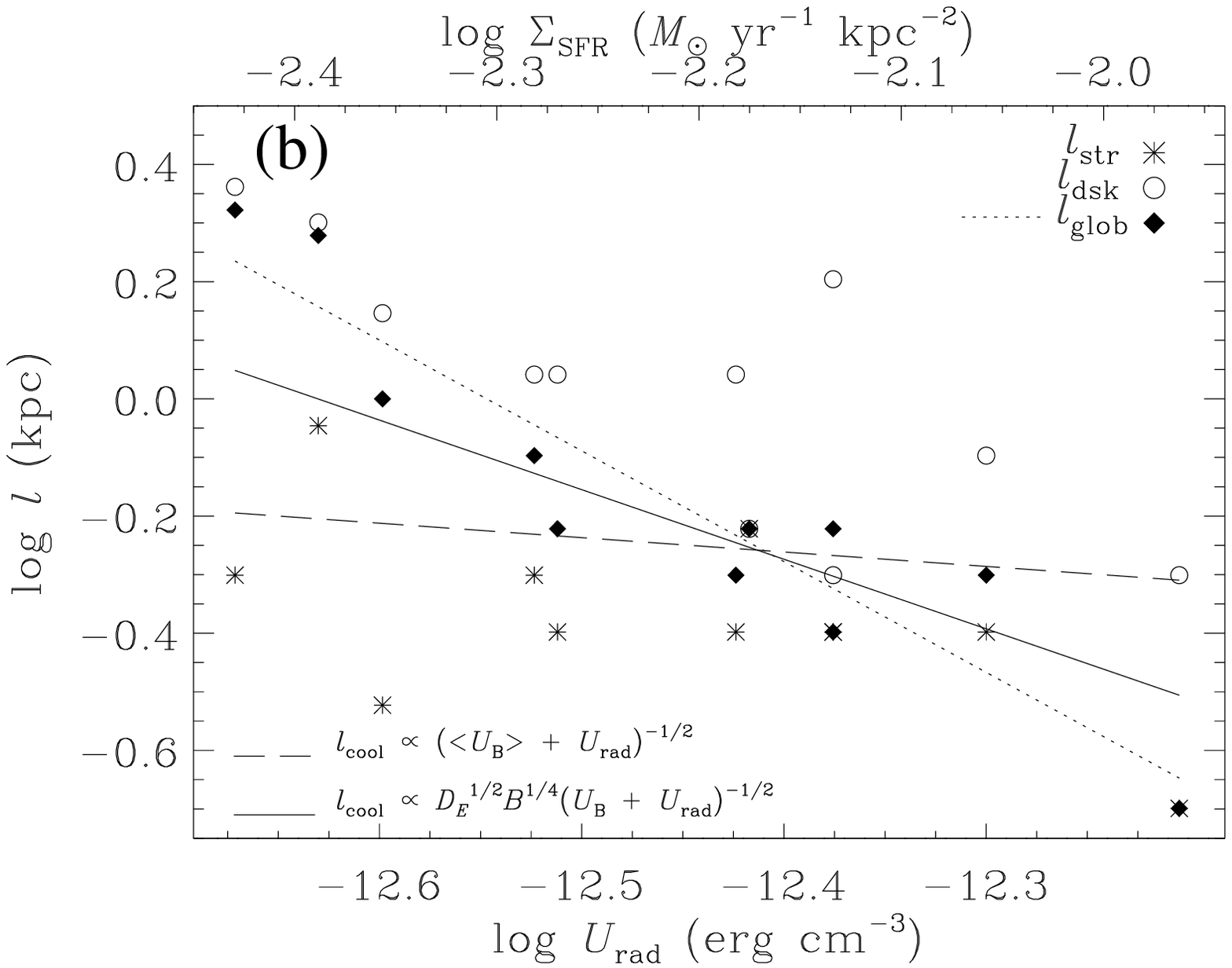}
  \caption{
    In panel (a) we plot the best-fit disk, structure, and global 
    scale-lengths for galaxies in our sample which are resolved at
    scales less than 1~kpc against the radiation field energy
    densities, ($U_{\rm rad}$, see $\S$\ref{sec-sfrd}).
    We exclude the galaxies Holmb~II, IC~2574, and NGC~4236 which
    have morphologies not well fit by our phenomenological
    model (see $\S$\ref{sec-resmaps}).
    Least square fits for the best-fit disk, structure, and global
    scale-lengths are plotted as {\it dot-dash}, {\it dashed}, and
    {\it dotted lines}, respectively. 
    In panel (b) we have excluded the galaxies NGC~925, NGC~4725,
    NGC~4559, and NGC~3031 (see $\S$\ref{sec-bfscl}).
    Along with the fit to the global scale-lengths ({\it dotted line})
    we also plot with the expected diffusion scale-lengths due to
    inverse Compton (IC) losses in a fixed magnetic field ({\it
    long-dashed line}) and synchrotron + IC losses with an
    energy-dependent diffusion coefficient $D_{E}$ for the steepest
    possible index ({\it solid line}; see $\S$\ref{sec-discCR}).   
    \label{SFRDles-gd}}
\end{figure*}

In simple diffusion models, the propagation of CR electrons is
usually characterized by an empirical, energy-dependent diffusion
coefficient, $D_{E}$ \citep[e.g.][]{ginz80}.
The value of $D_{E}$ has been found to be around \(4-6\times
10^{28}~{\rm cm^{2}~s^{-1}}\) for $\la$GeV CRs by fitting diffusion
models with direct measurements of CR nuclei 
(i.e. secondary-to-primary ratios like Boron-to-Carbon) 
within the Solar Neighborhood \citep[e.g.][]{fj01,im02,dm02}.
While this empirically measured value is for CR-nuclei within the
Milky Way, it has been found to be consistent with inferred diffusion
coefficients for CR electrons both radially
(\(\sim10^{29}~{\rm cm^{2}~s^{-1}}\)) and vertically 
(\(\sim10^{28}~{\rm cm^{2}~s^{-1}}\)) in the thin disks of galaxies
\citep[i.e. NGC~891 and NGC~4631;][]{dlg95}.  
We note that a value of \(\sim10^{29}~{\rm cm^{2}~s^{-1}}\) is also
found for hydrodynamic simulations of bubble/super-bubble induced
galaxy outflows in studies of galaxy cluster abundances
\citep[e.g.][]{pr05,er07}.
Since the mean model-derived diffusion coefficient, using direct
measurement of CR nuclei, is similar to the mean value of 
inferred radial and vertical CR electron diffusion coefficients by  
\citet{dlg95}, we simply use
\begin{equation}
\label{eq-DE}
\left(\frac{D_{E}}{\rm cm^{2}~s^{-1}}\right) \sim \Bigg\{
\begin{array}{cc}
5 \times 10^{28}, & E < 1~{\rm GeV}\\
5 \times 10^{28}(\frac{E}{{\rm GeV}})^{1/2}, & E \geq 1~{\rm GeV}.
\end{array}
\end{equation}

Now, neglecting escape and using a simple random-walk equation, we assume CR electrons will 
diffuse a distance \(l_{\rm cool} = (D_{E}\tau_{\rm cool})^{1/2}\)
before losing all of their energy to synchrotron and IC losses.
By combining Equations \ref{eq-nuBE} and \ref{eq-DE}, we can express
$D_{E}$ as a function of $B$ for a fixed $\nu_{\rm c}$ such that, for
CR electrons having energies $\geq$1~{\rm GeV},
\begin{equation} 
\label{eq-diff}
\left(\frac{l_{\rm cool}}{\rm kpc}\right) \sim 7 \times 10^{-4} 
\left(\frac{\tau_{\rm cool}}{\rm yr}\right)^{1/2}
\left(\frac{\nu_{\rm c}}{\rm GHz}\right)^{1/8}
\left(\frac{B}{\rm \mu G}\right)^{-1/8}.
\end{equation}

\subsubsection{Order-of-Magnitude Estimates \label{sec-oom}}
Using the above equations, we derive simple, order-of-magnitude estimates to determine whether diffusion and cooling of CR electrons in a steady-state star formation model are able to account for our observations.
In the following sections, more careful analyses are presented.
Taking the mean value of $U_{\rm rad}$ for those galaxies plotted in
Figure \ref{SFRDles-gd}b (i.e. $3.7\times 10^{-13}$erg~cm$^{-3}$), and
assuming $U_{\rm rad}=U_{\rm B}$, which has been shown to be a
reasonable assumption for a large sample of spiral galaxies
\citep{lvx96}, we find from Equation \ref{eq-cool} that the average
cooling time for a 1.4~GHz emitting CR electron is $\sim$$1.1\times
10^{8}$~yr.  
Inserting this value into Equation \ref{eq-diff}, we measure a
diffusion scale-length of $\sim$6.8~kpc; this value is clearly off of
the scale shown in Figure \ref{SFRDles-gd}b.  
On the other hand, if we instead assume a fixed, typical magnetic field 
strength of 9~$\mu$G \citep{sn95}, and that $U_{\rm B}=U_{\rm rad}=3.2\times
10^{-12}$~erg~cm$^{-3}$, the average cooling time for a 1.4~GHz emitting
CR electron is $\sim$$2.2\times 10^{7}$~yr with a diffusion
scale-length of $\sim$2.6~kpc.  
While this value for $U_{\rm rad}$ is much higher than what we infer
from the average TIR surface brightness of the sample, it must apply
near bright star-forming structures, whose TIR surface brightnesses
are much greater.  
Even so, this diffusion scale-length is much larger than any value we
find for the best-fit structure scale-lengths.
From these simple order of magnitude estimates it appears that
particle fading due to cooling by Inverse Compton and synchrotron
processes alone cannot explain the structural differences between the
FIR and radio maps.  

While CR electron escape may help to reconcile this discrepancy
between the observed scale-lengths and these order of magnitude
diffusion scale-length calculations, we might then expect to find
systematic variations in global FIR/radio ratios with best-fit
scale-lengths.
No such trend is found, suggesting that escape is not a
dominant variable among the galaxies included in Figure
\ref{SFRDles-gd}b.    
The role of escape, however, may be important for the sample
irregulars which is discussed in $\S$\ref {sec-discevo}.  
For the sample spirals, the best explanation seems to be differences in the CR electron population ages.

\subsubsection{Comparison with Scaling Relations}
We now introduce a more detailed model to see if we can reproduce
the observed trends in Figure \ref{SFRDles-gd}b by varying relevant ISM parameters using the above
mentioned scaling relations.
Again setting \(U_{\rm B} = U_{\rm rad}\) 
leads\footnote[5]{We note that this relation between $B$ and $U_{\rm rad}$ is slightly steeper than the $B \propto U_{\rm rad}^{1/3}$ scaling reported by \citet{nb97}.  If we were to instead use this scaling, the final relation in Equation \ref{eq-sclrel} would be less steep and even more discrepant with the fit to our observations in Figure \ref{SFRDles-gd}b.}
to \(B \propto U_{\rm rad}^{1/2}\) and \(\tau_{\rm cool} \propto B^{-3/2}\).
Since \(D_{E} \propto l_{\rm mfp}\) and \(l_{\rm mfp} \propto
n^{\delta}\), where $l_{\rm mfp}$ is the mean-free-path traveled by CR
electrons, $n$ is the ISM density, and $\delta$ depends on the
specifics of the scattering processes within a system, we 
can estimate the relative importance of $B$ and $n$ on the distance CR electrons 
emitting at a characteristic frequency travel. 
We introduce the scaling relations 
\begin{equation}
B \propto n^{\beta},~D_{E} \propto n^{\delta}B^{-1/4}, \nonumber
\end{equation}
where \(\onethird \leq \beta \leq \twothirds\) and \(-1 \leq \delta
\leq -\onethird\) are realistic index ranges \citep{hb93}, thereby
leading to the dependence
\begin{equation}
\label{eq-sclrel}
l_{\rm cool} \propto U_{\rm rad}^{1/4(\delta/\beta - 7/4)}.
\end{equation}

While the range of $\beta$ has been derived empirically
\citep[e.g.][]{nb97}, and is consistent with the assumption of
equipartition between turbulent kinetic energy of gas clouds and
magnetic energy, as well as with dynamo models \citep[e.g.][]{ruz88},
the values of $\delta$ are determined by the dominant CR electron
scattering processes.  
A value of $-1$ describes how diffusion would occur for hard
scattering off particles (i.e. the hard sphere approximation).
On the other hand, $\delta = -\onethird$ best describes
soft scattering off of structures; specifically, the case in which
scattering is dominated by localized $B$ perturbations whose spatial
density is proportional to the density of the gas \citep{hb93}.
In reality, both scattering processes are likely active which will
correspond to an intermediate value of $\delta$; 
this value will likely vary within and among galaxies.  

\begin{figure}[!ht]
  \plotone{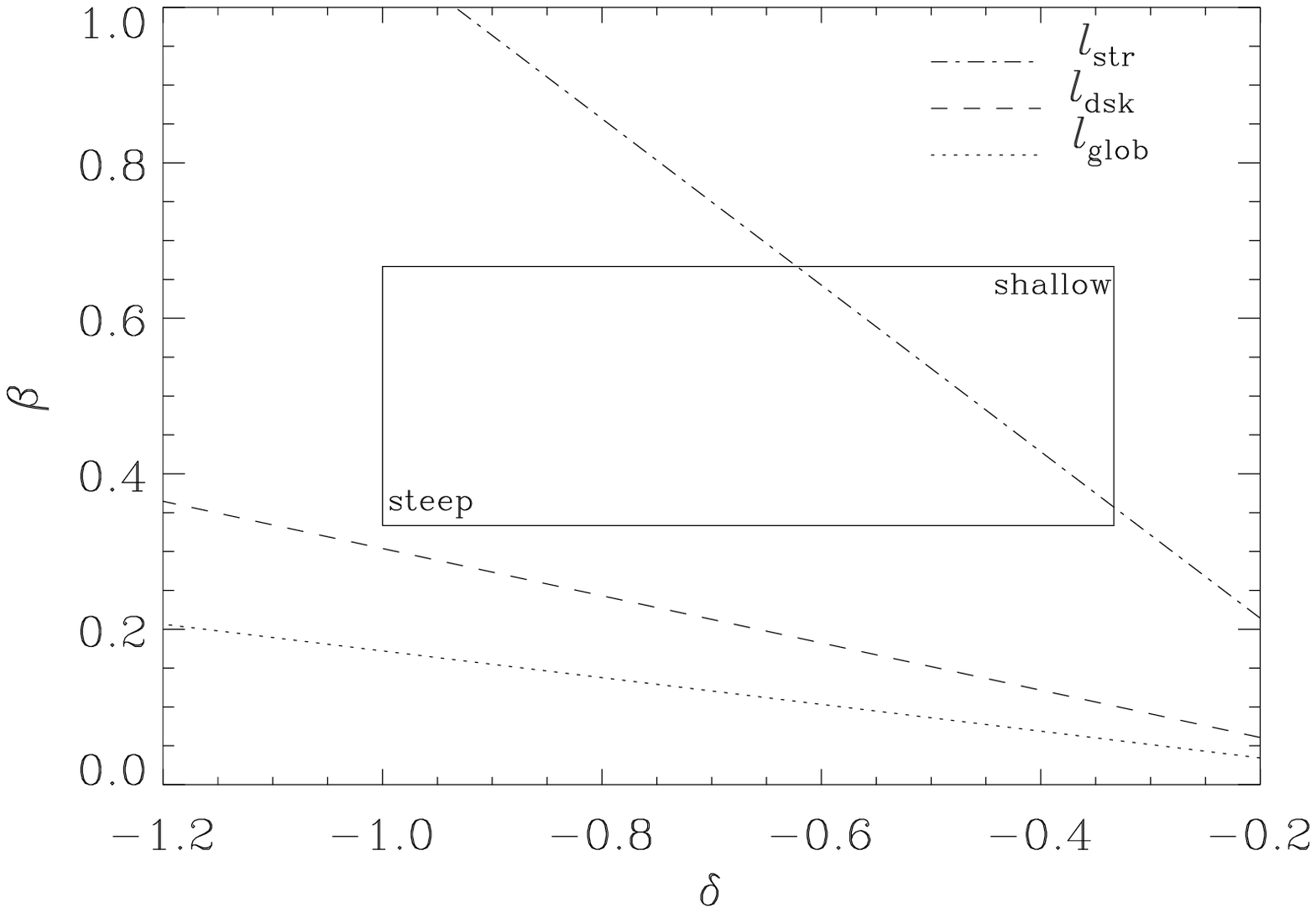}
  \caption{The slopes of the fits to global, structure, and disk
    scale-lengths as a function of $U_{\rm rad}$ (only for those galaxies shown in Figure \ref{SFRDles-gd}b), are plotted in terms of $\delta/\beta$ as given in Equation \ref{eq-sclrel}.   
    The overplotted box corresponds to the physically acceptable ranges
    of $\delta$ and $\beta$ (see $\S$\ref{sec-discCR}).
    \label{dbfit}}
\end{figure}

Excluding those galaxies which may suffer from low signal-to-noise
effects, we first take the simplest case and consider whether the 
observed trends in Figure \ref{SFRDles-gd}b may arise from an increase
of IC losses alone within galaxies that are more actively forming stars.  
Fixing $B$, we plot the expected relation between $l_{\rm cool}$ and
$U_{\rm rad}$ [i.e. $l_{\rm cool} \propto (<U_{\rm B}> + U_{\rm rad})^{-1/2}$] as a dashed line in Figure \ref{SFRDles-gd}b.
The fit to the global, structure, and disk data have slopes that are
$\sim$7.7, 2.7, and 5.1 times steeper than the slope of this line.
It is probably more physical for each parameter, not just $U_{\rm
  rad}$, to scale with the amount of star formation activity within
a galaxy; we will now investigate such a case.

Using the derived scaling relation given in Equation \ref{eq-sclrel},
and taking $\beta = \onethird$ and $\delta = -1$, we plot the expected
trend between $l_{\rm cool}$ and $U_{\rm rad}$ in Figure
\ref{SFRDles-gd}b as a solid line. 
These extreme choices of $\beta$ and $\delta$ correspond to the steepest possible slope and are unlikely to best describe most galaxies.
To more easily compare this model with the fits to the disk,
structure, and global scale-length data, we plot the slopes of each 
fit in terms of $\delta$ and $\beta$ (as given in the exponent of Equation \ref{eq-sclrel}) in Figure \ref{dbfit}. 
The slope of a given line in Figure \ref{dbfit} corresponds to a line
in Figure \ref{SFRDles-gd}b described by setting its slope equal to
the exponent in Equation \ref{eq-sclrel}.  
The physically acceptable ranges of $\delta$ and $\beta$ are given by
the overplotted rectangle.
The steepest and shallowest gradients correspond to coordinates of the
lower left and upper right corners of the rectangle, respectively.

Only the fit to the structure scale-length data passes through the
region of physically acceptable values for $\beta$ and $\delta$.
In particular, nearly the entire range of $\beta$ values, 
and values of $\delta \ga -\twothirds$, appear to be acceptable.
This suggests that diffusion occurring in star-forming
regions is best characterized as having a significant contribution from 
soft scattering off of perturbations in the magnetic field.
This scenario seems reasonable as the field near star-forming regions
is known to be amplified and highly turbulent \citep{rb96}.

The fit to the disk scale-lengths passes very near the plotted region
of physically acceptable scaling values in Figure \ref{dbfit}.
Specifically, values of $\beta = \onethird$ (weak magnetic field structure)
and $\delta = -1$ (hard scattering) appear the most relevant.

Finally, we find no physically acceptable combination of $\delta$ and
$\beta$ to reproduce the trend found between $l_{\rm glob}$ and
$U_{\rm rad}$. 
This result, in addition to the order of magnitude estimate in $\S$\ref{sec-oom}, suggests that systematic variations in ISM parameters
alone cannot properly describe the mean distance traveled by a
galaxy's CR electrons; 
rather, the age of the population must be the dominant parameter.

\subsection{Timescales of Star Formation Episodes}
Assuming that CR electron age is the dominant parameter which determines the measured best-fit scale-lengths, we can infer ages for the star formation episodes responsible for injecting the associated CR electron population into the ISM.  
We derive these ages from Equation \ref{eq-diff}, substituting $l_{\rm cool}$ values with the best-fit scale lengths.

The mean equipartition strength of the total magnetic field for typical
spiral galaxies is found to be $\sim$9~$\mu$G \citep{sn95}.
In strong spiral arms, such as those of NGC~5194, the total field
strength is found to be significantly larger at a value of
$\sim$30~$\mu$G \citep{af04}. 
The field strength in more quiescent regions of galaxies, like
that near the Sun, have typical values of $\sim$6~$\mu$G
\citep[e.g.][]{smr00}.  
Let us assume that each of these characteristic magnetic field
strengths correspond to the typical strengths for our global,
structure, and disk components, respectively, and scale with $U_{\rm
  rad}^{1/2}$ for the range in $U_{\rm rad}$ found for the
sample.  

The estimated ages for the global CR electron populations among the
sample galaxies range from $0.14 - 14. \times 10^{6}$~yr with a mean
age of $\sim$3.7~Myr.
The ages for the CR electron populations associated with the disk and
structure components range from $0.77 - 15. \times 10^{6}$~yr and
$0.19 - 3.4 \times 10^{6}$~yr, and have mean values of $\sim$5.3 and
1.2~Myr, respectively.  
Appropriately, we find the oldest ages to be larger than $10^{7}$~yr;   
this is approaching the average cooling time of a 1.4~GHz emitting
CR electron in a normal spiral galaxy.  
We also find that CR electrons residing in each galaxy's diffuse
disk are $\ga$4.4 times older, on average, than the population of
CR electrons associated with star-forming regions.

Furthermore, the galaxy exhibiting the highest star formation activity and shortest best-fit structure scale-length appears to have a dominant CR electron population with an age of $\sim2\times10^5$~yr. 
For this extreme case, such a young age suggests that the galaxy may still be in a phase of CR electron generation within the shells of SNRs; 
assuming an ambient particle density of 1~cm$^{-3}$ and an explosion energy of 10$^{51}$~erg, this age is $\sim$5 times larger than the adiabatic lifetime of SNRs, which by then should have expanded to have  a diameter on the order of $\sim$60~pc \citep[e.g.][]{jc92}.  

\subsection{Morphologies of Star Formation and CR Escape
  \label{sec-discevo}} 

\begin{figure}[!ht]
  \plotone{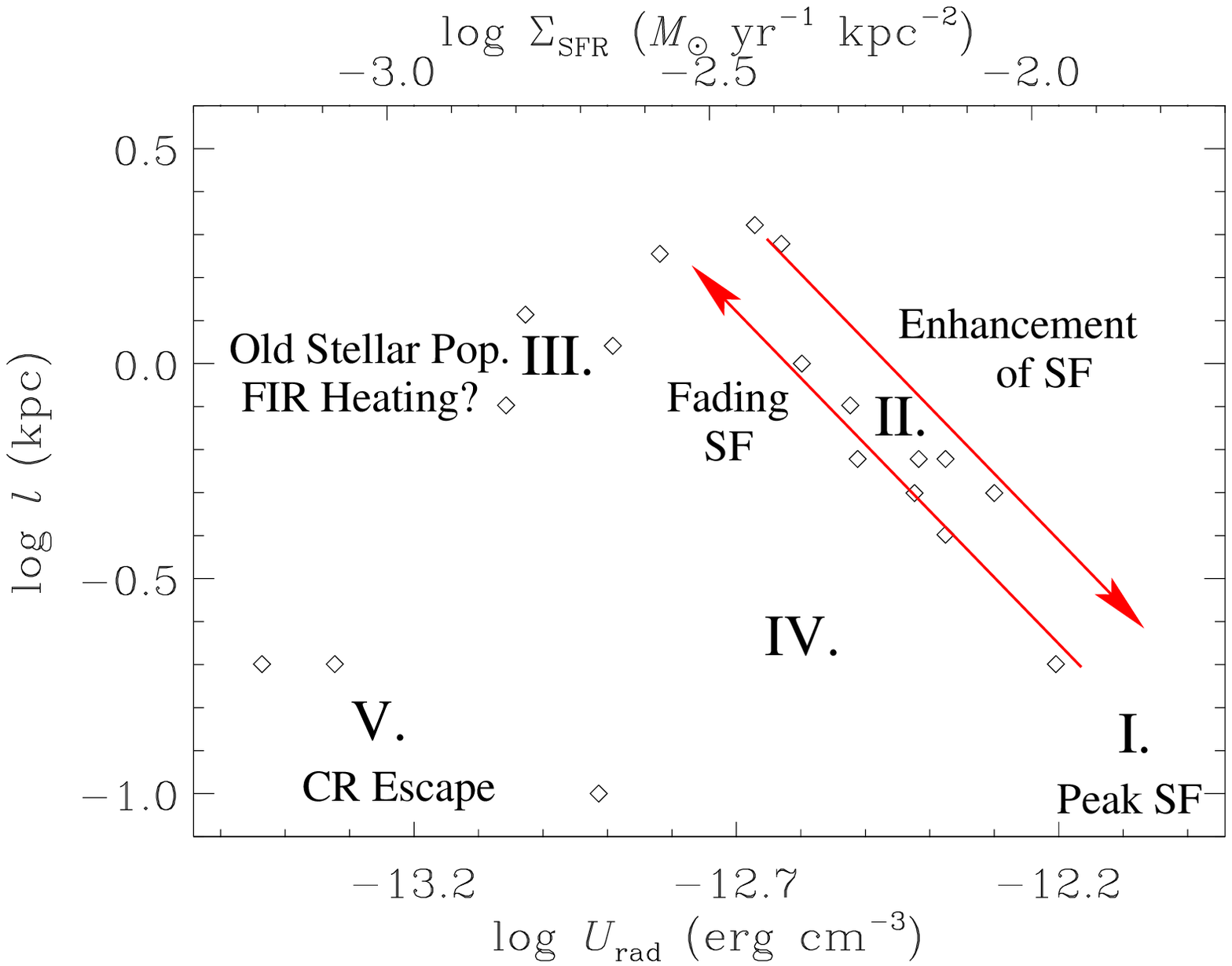}
\caption{The best-fit global scale-lengths for the entire 18 galaxy
  sample using isotropic kernels having an exponential profile and our
  free-free corrected radio maps.  
  Overplotted on the scatter diagram are roman numerals to identify
  the general placement of galaxies on this diagram (see
  $\S$\ref{sec-discevo} for the full discussion).
  \label{evol}}
\end{figure}

In order to provide a physical interpretation for the location of galaxies in the best-fit scale-length---$U_{\rm rad}$ diagram, we describe their general locations, designated by roman numerals, in Figure \ref{evol}.
So far in this discussion, we have not considered the four lowest
surface brightness galaxies (NGC~925, NGC~4725, NGC~4559, and
NGC~3031) because their placement in Figure \ref{SFRDles-gd}a may be
the result of a signal-to-noise effect.
Since it is difficult 
to determine whether this is in fact the case without having deeper
imaging, we will now speculate on a physical scenario which could
explain their positions in Figure \ref{SFRDles-gd}a.  
Similarly, the irregular galaxies in our sample (Holmb~II, IC~2574,
and NGC~4236) have also been excluded thus far in our 
discussion due to their lack of a disk component in the FIR and radio.  
The placement of these objects in Figure \ref{evol} and their behavior will now be put into the context of our phenomenological picture.

\begin{figure*}[!ht]
  \resizebox{18cm}{!}{
    \plotone{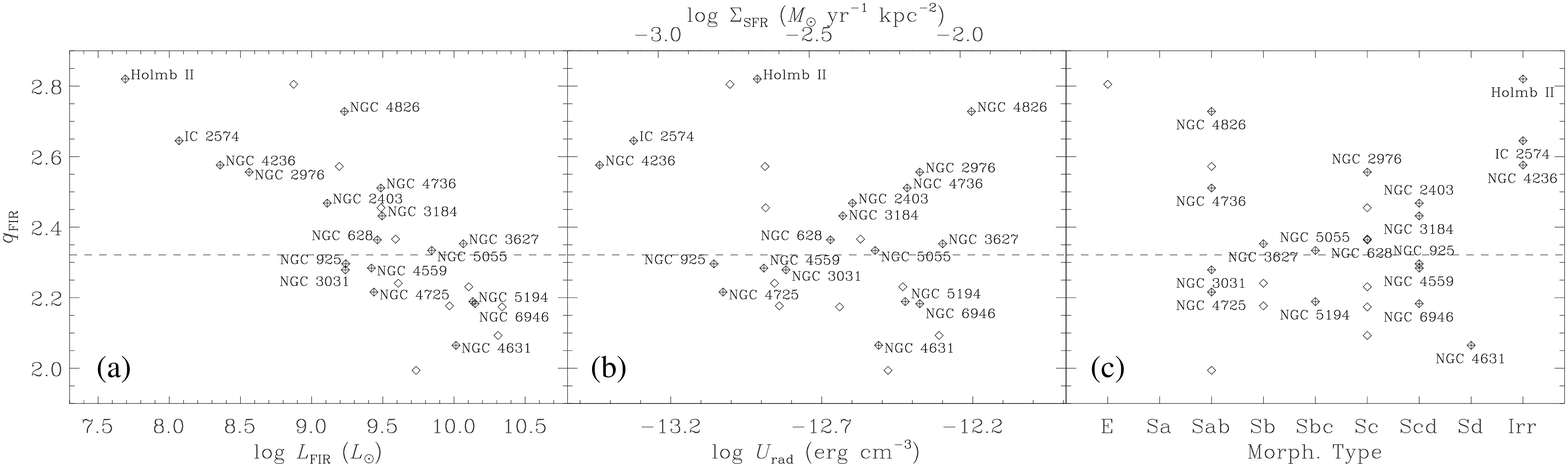}}
  \caption{
  We plot the global $q_{\rm FIR}$ ratio for the entire WSRT-SINGS
  sample as a function of FIR luminosity, radiation field energy
  density, and morphological type in panels a, b, and c,
  respectively.
  The horizontal line in each panel indicates the average $q_{\rm
  FIR}$ ratio for the entire sample. 
  Galaxies that are resolved at scales $\la 1$~kpc, and therefore
  included in the two-component image smearing analysis, have been
  labeled.  
  \label{qplot}}
\end{figure*}

\begin{enumerate}[I.]
\item
  {\it Peak of Most Recent Star Formation Event}:
  In this region of the diagram we assume a galaxy has just reached a
  peak in its surface brightness after a recent episode of enhanced
  star formation; 
  the episode must have occurred within the last few Myr to ensure that
  SNRs are still relatively young and that the morphologies of the 
  radio and infrared emission are similar because CR electrons have
  not had time to diffuse very far.
  While the nucleus of NGC~6946 has been classified as a mild
  starburst \citep{ball85}, we note that our sample does not contain
  any systems for which the entire galaxy disk is experiencing a
  strong starburst. 
  For systems experiencing such a mode of star formation,
  including the nuclear region of NGC~6946, high radiation fields and
  strong winds associated with the starburst could complicate
  this picture by rupturing magnetic field lines and increasing CR electron escape.   


\item
  {\it Enhancement and Fading of Star Formation}:
  After reaching a peak in global star formation activity, those CR
  electrons associated with the recent enhancement of star formation
  will dominate the CR population.
  As they spread through the galaxy they will begin to lose their
  energy to synchrotron and IC processes. 
  In $\S$\ref{sec-discCR} we have demonstrated that a galaxy's
  position along this trend cannot be due to variations in ISM parameters alone; 
  rather, its star formation history is the dominant parameter.
   
  Alternatively, galaxies may move along this part of phase space as
  star formation activity becomes enhanced and then fades.
  Assuming a mean $B$ of 10~$\mu$G and $U_{\rm rad} = U_{\rm B}$,
  1.4~GHz emitting CR electrons should lose their energy in
  $\sim$1.9$\times 10^{7}$~yr which, from Equation \ref{eq-diff}, 
  corresponds to a propagation length of $\sim$2.4~kpc.
  As this distance is consistent with the maximum values measured
  for our sample, we conclude that it takes a galaxy $\sim$5$\times
  10^{7}$~yr to completely traverse this region if it stopped forming
  stars after a single episode and all Type~II SNe occur on timescales
  $\la$3$\times10^{7}$~yr.  
  This is of course a lower limit, since galaxies have numerous star formation sites with a range of ages in their disks and star formation in spirals never completely turns off.  
   
  
\item  
  {\it FIR Heating by an Old Stellar Population?}:
  Galaxies in this region are characterized by low star formation activity and have best-fit scale-lengths which are shorter than expected given the trend found in region II.  
  While we expect the best-fit scale-lengths to reach some fixed value for galaxies having constant SFRs over timescales comparable to, or longer than, the CR electron cooling timescale,  and not increase indefinitely with decreasing infrared surface brightness, several possible explanations can be considered  for the observed turnover.

  Let us assume this is not a signal-to-noise effect.   
  One way to explain the shorter than expected scale-lengths is with
  an increase of CR electron escape.
  If this were the case we would then expect to find global FIR/radio
  ratios that are systematically higher than those for the more active
  star-forming galaxies.   
  Alternatively, this observation could be explained if the less
  active star-forming galaxies have larger thermal radio fractions
  than the higher surface brightness galaxies.
  In this scenario, we would instead expect to find global FIR/radio
  ratios which are systematically lower than what is observed for the
  galaxies with higher radiation field energy densities.  

  A non-linearity is in fact observed for the global FIR-radio
  correlation; 
  galaxies having $L_{\rm FIR} \la 10^{9} L_{\sun}$ have radio
  luminosities which are lower than what their FIR luminosities would
  predict \citep{jc92}.  
  In Figure \ref{qplot}a we plot the global $q_{\rm FIR}$ ratios
  versus FIR luminosity for the entire WSRT-SINGS sample.
  The known non-linearity is clearly observed.
  In Figure \ref{qplot}b we instead plot $q_{\rm FIR}$ against
  $U_{\rm rad}$; 
  here we do not see any evidence for the lowest surface brightness
  galaxies to exhibit systematically larger or lower FIR/radio ratios.
  It then seems that the above explanations (i.e. increased CR
  electron escape or larger thermal radio fractions)  are not able to
  reconcile the short scale-lengths found for objects in region III.  
 
  We also note that the trend found between the best-fit {\it
  structure} scale-lengths and radiation field energy density does not
  exhibit a turnover from regions II to III as seen for best-fit disk
  and global scale-lengths.  
  This suggests that the propagation of CR electrons associated with
  star-forming complexes is largely similar despite the differences in
  the galaxy's global star-formation activity.  
  Then, to simultaneously explain this result from our two-component
  analysis along with the depressed best-fit global scale-lengths, we
  may consider a situation in which older stars contribute
  significantly to the FIR heating.  
  This will make the FIR emission appear more diffuse and
  require less smoothing to best match the distribution of radio
  emission, as observed.

\item
  {\it Unpopulated Part of Phase Space}:
  We find no galaxies occupying this region of phase space in which relatively short global scale-lengths would be measured for a moderate values of $U_{\rm rad}$.  
  One way for a galaxy to populate this part of the diagram is if, after a long period of quiescence, it begins to form stars at a moderate rate.  
  The galaxy would then pass through this region very quickly ($\la$10$^{7}$~yr) before shifting into region II.  
  The lack of such galaxies suggests that, at least in the local Universe, star formation in spirals does not completely cease for long periods of time.


\item
  {\it CR Electron Escape}:
  As stated in $\S$\ref{sec-mscomp}, the irregulars behave markedly
  different than the sample spirals.  
  We do not believe this discrepant behavior to be the result of
  signal-to-noise effects.
  However, we do note that the radiation field energy densities of these galaxies, as derived from their infrared surface brightnesses alone, may be significantly underestimated due to the low dust content and more transparent media in irregular galaxies. 
  While shifting these objects to larger values of $U_{\rm rad}$ in
  Figure \ref{evol}, 
  they would still clearly remain outliers compared to the
  rest of the sample.
  The morphologies and ISM of these galaxies do not seem consistent with
  keeping their CR electrons bound outside of their initial clouds around SNRs.  
  After the CR electrons leave this cloud, the lack of a dense ISM and
  magnetic field to keep them trapped through multiple scatterings off
  of magneto-hydrodynamic (MHD) waves allows them to easily escape the
  system and enter intergalactic space. 

  Evidence for this picture comes from the extremely short best-fit
  scale-lengths measured for these galaxies.   
  The inability to retain CR electrons once they leave discrete
  star-forming regions should lead to deviations in the FIR-radio
  correlation;
  ratios should be significantly higher than the nominal value which
  is in fact observed.  
  In Figure \ref{qplot}c we plot the global FIR/radio ratio as a
  function of morphological type.
  All three irregular galaxies in our sample clearly exhibit FIR/radio
  ratios which are higher than the average value for the entire sample; 
  the average $q_{\rm FIR}$ for the irregulars is $\sim$0.34 dex
  larger than the mean for the entire sample.
  Furthermore, detailed multi-wavelength studies of the SINGS dwarf
  galaxies have uncovered these deviations {\it within}
  individual systems as well. 
  Strong variations ($\sim$1~dex) have been found in the TIR/radio
  ratio of IC~2574 \citep{jc05} while the radio continuum emission in
  the dwarf starburst NGC~1705 has been found to be an order of
  magnitude less than what one would expect from the nominal TIR/radio
  ratio \citep{jc06}.
  A more detailed, systematic study of the FIR and radio
  properties for dwarf galaxies is beyond the scope of this paper.
  
\end{enumerate}

\section{Remaining Issues and Future Prospects}
At the present, we appear to be reaching the limitations on what we can learn about the evolution of CR electrons in nearby galaxies with current data.  
Using more reliable SFR tracers \citep[e.g. 24~$\micron + {\rm H}\alpha$;][]{dc07,rk07} to compare with the radio should improve our ability to understand the effects discussed in $\S$\ref{sec-discevo}.  
Modeling is also needed and should lead to quantitative estimates for the SFR variations from the placement of a galaxy in Figure \ref{evol}; this can be a powerful tool for constraining the recent star formation histories of these galaxies. 

Due to our small sample size, especially for the lowest surface
brightness galaxies, a number of outstanding issues remain.
For one, it is currently difficult to determine whether the turnover
seen in Figure \ref{evol} is in fact physically driven or simply due to a signal-to-noise effect; deeper integrations are clearly needed.   
At the very high surface brightness end of the spectrum there is even less information to 
gauge how well our proposed phenomenology applies since the
present sample lacks any starbursting luminous infrared galaxies
(LIRGs); such galaxies are typically too distant to resolve on $\la$1~kpc scales with current FIR telescopes.
We speculate that such galaxies will behave similarly to the high surface
brightness galaxies in our sample, although strong galactic winds
arising from the central starburst could complicate the picture by
transporting CR electrons directly out of the disk, or perturbing
severely the magnetic field structure. 
The fact that such galaxies obey the FIR-radio correlation suggests
that the magnetic field must be amplified for synchrotron cooling
to retain the same fraction of CR electron energy losses relative to
escape and other competing processes. 
It is only with additional observations that these outstanding issues
can be addressed. 

Applying the smoothing phenomenology to such galaxies will add
important new insight on the role of magnetic fields and CRs in
the process of star formation and galaxy evolution as a whole;
starbursting ultra-luminous galaxies contribute significantly to the
luminosity density of the Universe during the most critical epochs $(z \ga1$) of
galaxy evolution.
More than 90\% of UV and visible radiation from starbursting
ultra-luminous LIRGs (ULIRGs) is absorbed by dust leaving FIR and
radio observations as the most promising way to gain insight on the
star formation processes within such systems and better understand the
physical connection between the gaseous and relativistic phases of the
ISM. 
Such a connection includes that of feedback processes;    
the production and expulsion of CRs in galaxies may work to cap their
luminosity and star formation intensity \citep{as07}.
The outlook for such additional observations appears rather bright;
with Herschel and the EVLA on the horizon, increasing the sample size
and improving the sensitivity and resolution of the observations
allowing the inclusion of more distant galaxies should become
achievable within a few years.

\section{Summary and Conclusions \label{sec-conc}}
Using a two-component image-smearing analysis, we have separated the
signatures of CR electron diffusion at spatial scales corresponding to
star-forming structures ($<$1~kpc) and galaxy disks ($\geq$1~kpc) 
within 18 galaxies observed as part of SINGS and WSRT-SINGS. 
Our results and conclusions can be summarized as follows:
\begin{enumerate}

\item
  We confirm and extend earlier results of M06a,b.  
  Empirically, the dispersion in the FIR-radio correlation within
  galaxies is most reduced by an image-smearing model; 
  this improvement is significantly better than what can be achieved
  by fitting correlations and removing linear trends.  
  
\item 
  The best-fit global scale-lengths decrease as a function of increasing star
  formation activity as measured by the infrared surface brightness of
  a galaxy.  
  Our interpretation is that a galaxy's CR electrons are closer
  to their place of origin within galaxies having intense star formation
  activity. 

\item
  The trend of decreasing best-fit {\it global} scale-length with
  increasing radiation field energy density is due to higher surface
  brightness galaxies having undergone a recent enhancement of star
  formation rather than variations in other ISM parameters.    
For sufficiently large enhancements, these galaxies are observed within $\sim$10$^{8}$~yr of the onset of the most recent star formation episode.

\item
  Unlike spirals, irregular galaxies 
  lack any well defined diffuse disk
  component at either 70~$\micron$, or especially at 22~cm.
  Presumably, the CR electrons escape these galaxies soon after
  leaving their parent star-forming regions due the absence of a
  dense ISM which would keep large-scale interstellar magnetic field
  locked into place.
  This conclusion helps to explain why these galaxies have global
  FIR/radio ratios systematically greater than canonical values.

\item 
As infrared surface brightness increases, the characteristic diffusion scale-length of a galaxy's CR electron population begins to transition at \(\log~U_{\rm rad} \leq -12.5\), or \(\log~\Sigma_{\rm SFR} \leq -2.3\), from being biased by CR electrons making up its diffuse disk to being biased by those recently injected near star-forming structures.
  From this we conclude that a galaxy's CR electron population 
  transitions from being dominated by old CR electrons to
  being dominated by young CR electrons as a function of star
  formation intensity.

\item
  The two-component analysis works better than smearing with a single
  smoothing kernel for spiral galaxies of type Sb or later which have
  high amounts of ongoing star formation activity (i.e. $\sim$40\% of the sample).
  This result suggests that star formation must be intense and highly structured for the two-component analysis of these data to differentiate properly between the different CR electron populations.


\end{enumerate}

\acknowledgements
We wish to thank the referee, Don Cox, for his insightful comments
which have greatly improved the paper.  
E.J.M is indebted to Eric Slezak for his highly constructive comments
and critical remarks that helped guide the technical aspects of the
wavelet decomposition.
E.J.M would also like to thank other members of the {\it Spitzer}
Infrared Nearby Galaxies Survey (SINGS) team for their invaluable
contributions to the work presented here, especially those of
R.C.Kennicutt, Jr., D.Calzetti, K.D.Gordon, C.W.Engelbracht, and
G.Bendo. 
As part of the {\it Spitzer} Space Telescope Legacy Science Program,
support was provided by NASA through Contract Number 1224769 issued by
the Jet Propulsion Laboratory, California Institute of Technology
under NASA contract 1407.  
E.J.M. also acknowledges support for this work provided by the {\it
  Spitzer} Science Center Visiting Graduate Student program.
This research has made use of the NASA/IPAC Extragalactic Database
which is operated by JPL/Caltech, under contract with NASA. 
This publication makes use of data products from the Two Micron All
Sky Survey, which is a joint project of the University of
Massachusetts and IPAC/Caltech, funded by NASA and the National
Science Foundation.

\end{document}